# Design of Sliding Mode PID Controller with Improved reaching laws for Nonlinear Systems

*Submitted in partial fulfillment of the requirements for the degree of*

**Master of Technology**

*by*

**Kirtiman Singh**

**Roll No. 1220502**

*Supervisor:*

**Dr. Prabin Kumar Padhy**

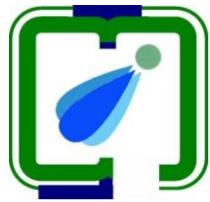

**MECHATRONICS**

**PDPM INDIAN INSTITUTE OF INFORMATION TECHNOLOGY, DESIGN AND MANUFACTURING JABALPUR**

**(May, 2014)**

To See the World,

Things dangerous to come,

To see behind walls,

To draw closer,

To find each other and to feel,

That is the purpose of **LIFE.**

**Dedicated to my Guide, Parents and my Brother**

# Approval

Thesis entitled "**Design of Sliding Mode PID Controller with improved reaching laws for Non-Linear System**", by **Kirtiman Singh, Roll No. 1220502** is approved for the degree of Master of Technology.

\_\_\_\_\_\_\_\_\_\_\_\_\_\_\_\_\_\_\_\_\_

\_\_\_\_\_\_\_\_\_\_\_\_\_\_\_\_\_\_\_\_\_

\_\_\_\_\_\_\_\_\_\_\_\_\_\_\_\_\_\_\_\_\_

**Examiners**

\_\_\_\_\_\_\_\_\_\_\_\_\_\_\_\_\_\_\_\_\_

**Supervisor**

\_\_\_\_\_\_\_\_\_\_\_\_\_\_\_\_\_\_\_\_\_

**Chairman**

Date: \_\_\_\_\_\_\_\_\_\_

Place: \_\_\_\_\_\_\_\_\_\_

# Certificate

This is to certify that the thesis entitled, "**Design of Sliding Mode PID Controller with improved reaching laws for Non-Linear System**", submitted by **Kirtiman Singh, Roll No. 1220502** in partial fulfillment of the requirements for the award of Master of Technology Degree in Mechatronics Engineering at the PDPM Indian Institute of Information Technology, Design and Manufacturing Jabalpur. This is an authentic work carried out under my supervision as well as guidance and this work has not been submitted elsewhere for degree.

To the best of my knowledge, the matter embodied in the thesis has not been submitted elsewhere to any other university/institute for the award of any other Degree.

**Dr. Prabin Kumar Padhy**
(Associate Professor)
Electronics and Communication Engineering
PDPM Indian Institute of Information Technology,
Design and Manufacturing Jabalpur (M.P.),
482005, INDIA
Email: prabin16@iiitdmj.ac.in

Date: __________

Place: __________

# Acknowledgement

Behind this thesis dissertation lies a year of hard work. It is therefore a great pleasure for me to have an opportunity to thank all those who had given me the support and encouragement to write this thesis. First of all, I would like to express my profound gratitude to my thesis advisor **Dr. Prabin Kumar Padhy** (Department of Electronics & Communication, PDPM IIITDM Jabalpur) for his guidance during my thesis and study at PDPM IIITDM. From finding an appropriate area of interest in the beginning to the process of completion, Dr. Prabin Kumar Padhy offered his unreserved help and guidance and led me to finish my thesis with a lot of freedom in conducting the research. He was always accessible and willing to help. His wisdom and expert guidance have been indispensable in the elaboration of this thesis and without his efforts; I would not have the support to finish this work.

A special thanks to my parents and brother, for their continuous love, patience, support and encouragement in my decisions. Without whom I could not have made it here.

My final thanks go to **Mrs. Swapnil Nema** who squeezed her time from her busy schedule and motivated and helped me from time to time.

.

Date:

Kirtiman Singh

# Abstract

In this thesis, advanced design technique in sliding mode control (SMC) is presented. This thesis focuses on PID (Proportional-Integral-Derivative) type Sliding surfaces based Sliding mode control with improved power rate exponential reaching law for Non-linear systems using Modified Particle Swarm Optimization (MPSO) [4]. Non-linear dynamical systems are of interest to engineers, physicists and mathematicians because all real physical systems are inherently non-linear in nature. It is advantageous to consider these non-linearities directly while designing controllers for such system. Besides the ubiquitous PID controllers, a lot of research effort has been put on robust and nonlinear controllers, which is a model based technique. This method is then evaluated by studying their model for the particular choices of non-linear systems. They are on path to become the main tools in advance aerospace and industrial feedback loops and are rapidly advancing to their new versions. So in order to handle large non-linearities directly, sliding mode controller based on PID-type sliding surface has been designed in this work, where Integral term ensures fast finite convergence time. The controller parameter for various modified structures can be estimated using Modified PSO, which is used as an offline optimization technique. Then various reaching law were implemented leading to the proposed improved exponential power rate reaching law, which also improves the finite convergence time. To implement the proposed algorithm, nonlinear mathematical model has to be decrypted without linearizing, and used for the simulation purposes. Their performance is studied using simulations to prove the proposed behavior. The problem of chattering has been overcome by using boundary method and also second order sliding mode method. PI-type sliding surface based second order sliding mode controller with PD surface based SMC compensation is also proposed and implemented. The above proposed algorithms have been analyzed using Lyapunov stability criteria. The robustness of the method is provided using simulation results including disturbance and ±10% variation in system parameters. Finally process control based hardware is implemented (conical tank system).

# Table of Contents

# List of Figures

# List of Tables

# Chapter 1

# Introduction

## 1.1 Overview

Non-linear dynamical systems are considered as an interesting topic of investigation as all real systems are always nonlinear and linearizing them can impose too strict requirements on their working range or the feasible results. So, nonlinearity can sometimes be intently introduced in the feedback control loop as nonlinear systems can provide better performance than the linear ones and can accurately control the systems behavior,. The presence of uncertainties and disturbances can further add complexity to the control of real dynamical systems as it is impossible to avoid them due to modeling uncertainties and environmental effect. The effects of these nonlinearities, uncertainties and disturbances have to be clearly taken into account since they can worsen the performance or even cause instability. Non-linearity is a common behavior exists in all the processes and it disturbs in the performance of the system. Special attention is required to design a controller for non-linear system. The non-linearity can be understood through the following forms like Static Non-linearity, Dynamic Non-linearity, Intentional Non-linearity and Incidental Non-linearity. The behavior of the non-linear system can also be explained by following properties

- Multiple equilibrium points.
- Limit cycles: Oscillations of fixed amplitude and fixed period without external excitation.
- Sub harmonic, harmonic oscillations for constant frequency inputs.
- Chaos:
    - randomness, complicated steady state behaviors.
    - System output is extremely sensitive to initial conditions.
    - Unpredictability of the system output.

The control laws governing dynamical system need to have nonlinear and time varying behaviour with various uncertainties, disturbances and parasitic dynamics, the influence of which has to be carefully taken into account when considering the performance of the system [1-2]. System model imprecision or discrepancies may come from actual uncertainty about the plant or from the purposeful choice of a simplified representation of the system's dynamics. These mismatches may be due to various factors. These modelling inaccuracies can have strong adverse effects on control systems. The principle objective of control law is to generate a control or regulation system which is robust to external disturbances and model uncertainty. The control system has seen development ranging from the ubiquitous PID controller through to high performance controllers [4]. PID controllers are one of the most popular and simple controller used by the engineers mainly in process industries, automatic control etc. Adding to the value of PID's are optimal and robust controllers [10-18]. These control algorithms are independent of the model, linear in nature and their parameter has to be adjusted to obtain an acceptable response characteristic. Recently, there has been a lot of activity in the design of the so-called intelligent control techniques such as fuzzy logic and neural network [5-9], which rely on learning the input-output behavior of the plant to be controlled. Superseding them are the mature mathematically model based controllers in the area of nonlinear control which yield high performance, near world tests of its classes in the aerospace industry. Instead of directly applying the intelligent techniques, their role has been reversed to processes which are ill defined, nonlinear, complex, time varying and stochastic [17]. One of the most important approaches for dealing with model uncertainty is robust control [12]. The typical structure of a robust controller is composed of a nominal part, similar to a feedback control law, and additional terms aimed at dealing with model uncertainty. Also, nonlinear behaviour is sometimes introduced in feedback control as they can provide for even better performance than the linear ones. A key reason for using feedback is to reduce the effects of uncertainty which may appear in different forms as disturbances or as other imperfections in the models used to design the feedback law. Model uncertainty and robustness have been a central theme in the development of the field of automatic control. Controlling a nonlinear dynamical system in presence of heavy uncertainty conditions is the mainstay of this thesis. Considerable advancements have been made in robust control techniques such as adaptive control, model predictive control, H-infinity control and backstepping. However these control approaches can only deal with linear system or the linearized models of the nonlinear systems. A more advanced solution for the nonlinear system is Sliding-Mode Control (SMC) since it is simpler than other controllers like H-infinity and it can deal with both uncertain linear and nonlinear systems. Sliding mode control is the principle operation mode of VSC, characterized by high simplicity and robustness [44-46]. The concept of variable structure control (VSC) was proposed and elaborated in the early 1950's by Emelyanov [30-32] in Soviet Union, considering only linear second-order systems. Since then, VSC has developed into a general design method for nonlinear systems, multi-input multi-output systems, discrete time models and stochastic systems [34-36]. The most distinguishing feature of VSC is its ability to result in very robust or invariant control system. Loosely speaking, the term 'invariant' means that the system is completely insensitive to parametric uncertainties and external disturbances. "Variable structure control systems" refers to a class of systems whereby the "control law" is deliberately changed during the control process according to some defined rules which depend on the state of the system [35, 37-38]. It utilizes discontinuous control law which drives the systems state trajectory onto the sliding or switching surface in the state space and to keep the system state onto the sliding manifold for all subsequent times. Therefore, control input is designed to overcome the uncertainties and disturbances acting on real dynamical systems. Main advantages of sliding mode control are its behavior as a reduced order plant when on the sliding manifold, and its insensitivity towards model uncertainties and disturbances [47-48]. Besides the robustness properties, the real-life implementation of SMC techniques presents a unique problem of 'Chattering' i.e. high-frequency oscillations of the controlled system [59]. This phenomenon occurs due to the inertias of the actuator and sensors, in the presence of noise and external disturbances, only switches its state at a finite frequency contrary to the control signal which switches at infinite frequency [60]. Therefore, Chattering and discontinuous control action are the two main drawbacks of sliding mode control techniques. It is also limit the behavior according to the nature of disturbance acting on the system. To overcome or attenuate the above cited drawbacks, this work analyzes recent developments happening in sliding mode control, namely boundary condition method and second order sliding mode control [26]. Boundary approaches attenuates the chattering behavior of the system while confining the system response along the boundary width, implemented using various functions like 'saturation' function, 'tanh' function etc. Second order sliding mode control produces a continuous control output while keeping the advantages of the original approach, and provide for even higher accuracy in realization.

The objective of this thesis is to survey the theoretical backgrounds of sliding mode control. Sliding mode control comprise of two steps: Sliding surface design and subsequently control law design for the sliding hyper plane manifold. In this thesis, an integral term is included in the sliding surface definition that results in a PID-type sliding surface [39-41]. This thesis discusses and investigates the advantages of such surface design using simulations. Proceeding on that, PI-like sliding surface is used for second order sliding mode control with compensation schemes to exploit its advantages and provides an effective solution to the above cited drawbacks, providing some original contribution to the thesis. Higher order sliding mode control was discussed by Levant in the 1987 in his Phd work. In normal sliding mode, the sliding variable is kept at the origin but its derivative may be non-zero. For higher order sliding mode, first and the higher order derivatives of the sliding variable are also forced to zero during sliding mode, meaning higher precision [57-58]. With a higher order sliding mode controller, the amplitude of the chattering effect can be dramatically reduced. Coming to the second part of sliding mode i.e. control law design, implemented using equivalent control law and reaching control laws [35]. Reaching control laws are then modified for a better response of sliding mode control design and are presented in the thesis with simulations and mathematical analysis. In particular, Sliding mode control methodology using PI-type sliding surface is applied to second order systems. Apart from the robustness features against different kind of uncertainties and disturbances, the proposed control schemes have the advantage of producing low complexity control laws compared to other robust control approaches (H∞, LMI, adaptive control, etc.) which appears particularly suitable in the considered contexts.

## 1.2 Literature Survey

Sliding mode control has its roots in (continuous-time) relay control. It originated in the Soviet Union somewhere in the late 1950s [30-31], but was not published outside the Soviet Union until the publication by Utkin [34]. After this publication, the list of publications concerning SMC grew rapidly and SMC obtained a solid position in the field of both linear and nonlinear control theory. To obtain a good overview of continuous-time sliding mode we refer to the standard work of Utkin [44]. Various other good publications are DeCarlo, Hung, Edwards respectively [35, 36, 45,], where the latter publications describe several practical applications. Sliding mode control for nonlinear systems are reported by Slotine, and by Lu and Spurgeon [42, 45]. Chattering elimination is a hot topic for sliding mode control design. Significant contribution has been made by Bartolini, Levant and Fridman [59-61]. Output feedback is recognized one of the effective approaches for tracking error elimination. It has been combined with sliding mode control [75]. Besides the continuous sliding mode control scheme, discrete time sliding mode has been deeply investigated. Systems with time delay are treated with sliding mode control [74]. In recent years, a so-called "higher order sliding mode (HOSM)" has been proposed by Emelyanov, Levant and Gao, for nonlinear sliding mode design [32, 57-58]. So-called "twisting controller" and "super-twisting controller" are regarded as the most popular types of higher order sliding mode controllers [46]. The structure of the controller and parameter selection rules of such controllers is developed and the benefit of this second or higher order actuator is detailed in the reference. It is concluded that there is opportunity for reduction of the amplitude of chattering in the control signal when using twisting as a filter algorithm, compared with first order sliding mode control. Levant has also proposed a higher order sliding mode observer design [46].

## 1.3 Research Objective

The non-linear systems are very hard to analyze and control explicitly. However, qualitative and numerical techniques may help in describing some information on the behavior of the solutions. It is always advantageous to consider the non-linearities directly while analyzing and designing controllers for such systems. Designing of a Sliding mode controller based on a PID-type sliding surface is mainly aimed to control the nonlinear dynamic systems. To tune the three parameters, which are proportional ($K_p$), integral ($K_i$) and derivative ($K_d$) of the controller are approximated using Modified PSO. This can be a solution for non-linear dynamical systems which is the designing of sliding mode controller directly using PID-type sliding surface, whose advantages are:

- no approximation approach
- handles the non-linear system directly
- can handle the nonlinearities in large range operation as well
- very high convergence rate

The overall objective of this thesis is to develop a sliding mode controller structure based on PID-type sliding surfaces whose parameters are tuned by modified PSO algorithm forming a solution to non-linear dynamical systems.

1. Justification of traditional sliding mode controllers towards non-linear dynamical systems.
2. A PID-type sliding surface based sliding mode controller structure with the concept of internal feedback and parameters tuned using modified PSO algorithm.
3. A novel controller structure with PI-type sliding surface based second order sliding mode controller in feed forward path along with PD-type sliding surface based sliding mode controller in feedback path with stability analysis.
4. Finally, the implementation of the controller structure was performed on SimMechanics model of an Inverted Pendulum and others like conical tank and van der pol's equation.

## 1.4 Thesis Outline

This thesis is organized in *six chapters*.

*Chapter one* gives an introduction about the thesis work including the literature survey. The detail design method of Sliding mode control for a class of non-linear systems is given in *Chapter two*. It describes about the basics of the sliding mode control design. It also details about the sliding manifold design, control law design, drawbacks of the sliding mode control design, modified particle swarm optimization, and also about the system's physics which plays a direct role on the design of the control law design. *Chapter three* describes a modified PID-type sliding surface based sliding mode controller structure in which an additional integral term is used in the controller parameters. All the controller parameters are estimated using PSO and different non-linear processes have been considered for the validation of the proposed controller structure. *Chapter four* describes a novel controller structure with a proportional integral (PI) type sliding surface based sliding mode controller in the feed forward path and a proportional derivative (PD) type traditional sliding mode controller acting in the feedback path. The stability of the system is ensured by the PD type sliding mode controller in the proposed structure and proved by direct Lyapunov stability criteria. The robustness of the method is illustrated by simulation in MATLAB using SimMechanics model of the system. *Chapter five* is about the hardware implementation. Non-linear system (conical tank) is analyzed with considerable mathematical modeling and the performance is verified using proposed PID-type Sliding mode controller. The water level of the conical tank is maintained at a level. *Chapter six* summarizes the concluding remarks of the thesis work and the future scope.

# Chapter 2

# Sliding Mode Control

## 2.1 Introduction

Sliding mode control (SMC) is a special case of variable structure control [35]. Sliding mode control methodology is based on the remark that it is much easier to control a first order system (i.e. systems described by first order differential equation), be they nonlinear or uncertain, than to control a general $n^{th}$ order system (i.e. systems described by $n^{th}$ order differential equation) [19]. Sliding mode control is applicable to nonlinear systems as it directly considers robustness issue as a part of the design process [42]. It is a nonlinear control method that alters the dynamics of a nonlinear system by application of a high frequency switching control [43]. The state-feedback control law is not a continuous function of time. Instead, it switches from one continuous structure to another based on the current position in the state space. Hence, sliding mode control is a variable structure control method [44-46]. The multiple control structures are designed so that trajectories always move toward a switching condition, and so the ultimate trajectory will not exist entirely within one control structure. Instead, the ultimate trajectory will slide along the boundaries of the control structures [47-53]. The motion of the system as it slides along these boundaries is called a sliding mode and the geometrical locus consisting of the boundaries is called the sliding (hyper) surface. The sliding surface is described by $s = 0$, and the sliding mode along the surface commences after the finite time when system trajectories have reached the surface [56]. Intuitively, sliding mode control uses practically infinite gain to force the trajectories of the dynamic system to slide along the restricted sliding mode subspace. Trajectories from this reduced-order sliding mode have desirable properties. The system dynamic when confined to the sliding surface is described as an ideal sliding motion and represents the controlled system behaviour. The advantages of obtaining such a motion are two-fold: Firstly, the system behaves as a system of reduced order with respect to the original plant; and secondly the movement on the sliding surface of the system is insensitive to a particular kind of perturbation and model uncertainties. The main strength of sliding mode control is its robustness. Because the control can be as simple as a switching between two states, it need not be precise and will not be sensitive to parameter variations that enter into the control channel. Additionally, because the control law is not a continuous function, the sliding mode can be reached in finite time (better than the asymptotic behaviour). This property of invariance towards so called matched uncertainties or arbitrary parameter inaccuracies is the most distinguish feature of sliding mode control and makes this methodology particular suitable to deal with uncertain nonlinear systems [50]. Such performance however comes at a price of extremely high control activity. For general nonlinear systems, sliding mode controller provides a systematic approach to the problem of maintaining stability and consistent performance in the face of modelling imprecision.

## 2.2 Problem Formulation

Consider the following nonlinear system affine in the control

$$\dot{x} = f(x,t) + g(x,t).u(t) \qquad (2.1)$$

where, $x(t) \in R^n, u(t) \in R^m$, and

$f(x,t) \in R^{n*n}$ is not exactly known but the extent of imprecision is upper bounded by a known continuous function of $x$, and

$g(x,t) \in R^{n*m}$ is not exactly known, but is of known sign and is bounded by known, continuous function of $x$.

The component of the discontinuous feedback is given by:

$$u_i = \begin{cases} u_i^+(t,x) & if \quad s_i(x) > 0 \\ u_i^-(t,x) & if \quad s_i(x) < 0 \end{cases} \quad if \quad i=1,2,......,m. \qquad (2.2)$$

where, $s_i(x) = 0$ is the $i - th$ sliding surface, and

$$s(x) = [s_1(x), s_2(x), \ldots, s_m(x)]^T = 0 \tag{2.3}$$

is (n-m) dimensional sliding manifold in $R^n$, determined by the intersection of the $m(n-1)$ sliding manifold. Since $u_i(x,t)$ undergoes discontinuity on the surface $s_i(x) = 0$ ($s_i(x) = 0$ is called switching surface or switching hyperplane), the switching surface is designed such that the system response restricted to the surface equation has the desired behaviour.

After switching surface design, the next important aspect is guaranteeing the existence of a sliding mode. Let $S = \{x | s(x) = 0\}$ be the switching surface that can be scalar, linear, stable ordinary differential equation (nonlinear) which includes the origin $x = 0$, then $x(t)$ is the sliding mode of the system. A sliding mode exists, if in the vicinity of the switching surface, the tangent or velocity vectors (time derivative of the state vector) of the state trajectory always point towards the switching surface. Switching surface $S$ is called sliding surface or sliding manifold if for every point on the surface there are trajectories reaching it from both sides of the surface. To prove the existence of the sliding mode, stability of the state trajectories in the neighbourhood of the sliding surface $S$ is required i.e. representative point must approach the surface at least asymptotically. The sufficient condition for the sliding mode is called reaching condition, termed as Reaching mode and the largest neighbourhood which satisfies this reaching condition is called the region of attraction. The existence problem can be seen as a generalized stability problem, using second Lyapunov criteria. Specifically, stability requires the candidate Lyapunov function which is positive definite, to have a negative time derivative in the region of attraction. The control problem is to get the state $x$ to track a specific time varying state $x_d = [x_d, \dot{x}_d, \ldots, x_d^{(n-1)}]$ in the presence of model imprecision on $f(x)$ and $g(x)$.

Design of Sliding mode control consists of two parts:
1. Design of the sliding surface, usually of order $(n-1)$ or smaller, must be constructed such that the system performance satisfies the design objectives like stability, order reduction etc.
2. Switching feedback control is designed such that the reaching condition is satisfied which drives the state trajectory to the sliding surface in finite time and maintains it thereafter.

## 2.3 SLIDING SURFACE DESIGN

Sliding surface can be linear or nonlinear. Linear switching surface for linear dynamic has been developed in great depth and completeness whereas for nonlinear surfaces, design of switching surface remains largely an open problem.

Consider a general second order nonlinear system,
$$\ddot{x} = f(x,t) + g(x,t).u(t) \tag{2.4}$$
with a sliding surface $S = \{x|s(x) = 0\}$
where, $x(t) \epsilon\ R^n$, $u(t) \epsilon\ R^m$, $f(x,t) \epsilon\ R^{n*n}$, and $g(x,t) \epsilon\ R^{n*m}$.

### 2.3.1 Time varying surface

Let the tracking error vector be defined as
$$e = x - x_d = [e\ \dot{e}\ \ldots\ldots\ e^{(n-1)}] \tag{2.5}$$

For a single input system, time varying sliding surface in the state space $R^n$ is defined by the scalar equation $s(x,t) = 0$ according to the desired control bandwidth
$$s(x,t) = \left(\frac{d}{dt} + \lambda\right)^{(n-1)}.e = 0 \tag{2.6}$$
where, $\lambda$ is a strictly positive constant which determines the closed loop bandwidth. For example, if n=2
$$s = \dot{e} + \lambda e \tag{2.7}$$
i.e. $s$ is simply the weighted sum of the position error and the velocity error; if n=3
$$s = \ddot{e} + 2\lambda\dot{e} + \lambda^2 e \tag{2.8}$$

Given the initial condition $x_d(0) = x(0)$, the problem of tracking $x = x_d$ is equivalent to that of remaining on the surface $s(t)$ for all $t > 0$ i.e. the original n[th] order tracking problem in $x$ in effect be replaced by a first order stabilization problem in $s$. It can also be seen that the scalar $s$ represents a true measure of tracking performance as bounds on $s$ can be directly translated into bounds on the tracking error vector representing a true measure of tracking performance.

## 2.4 CONTROL LAW DESIGN

The systems motion on the sliding surface can be given as an interesting geometric interpretation, as an average of the system's dynamics on both sides of the surface. Upon the interjection on the sliding surface, problem of reachability has to be solved. It involves selection of a state feedback control function $u: R^n \to R^m$ which forces the state $x$ towards the sliding surface and thereafter maintains it on the surface i.e. the controlled system must satisfy the reaching condition. The dynamics while in sliding mode can be written as $\dot{s}(x) = 0$. By solving the equation formally for the control input, obtained is an expression for $u$.

### 2.4.1 Equivalent Control method

Equivalent control is obtained by recognizing $\dot{s}(x) = 0$, which is the necessary condition for the state trajectory to stay on the sliding surface $s(x) = 0$ if the dynamics is exactly known. Therefore, setting $\dot{s}(x) = 0$, i.e.,

$$\dot{s} = \frac{\partial s}{\partial x} \cdot \dot{x} = \frac{\partial s}{\partial x} \cdot f(x) + \frac{\partial s}{\partial x} \cdot g(x) \cdot u_{eq} = 0 \qquad (2.9)$$

Solving for $u_{eq}$ yield the equivalent control,

$$u_{eq} = -\frac{\partial s}{\partial x} \cdot f(x) \cdot \left(\frac{\partial s}{\partial x} \cdot g(x)\right)^{-1} \qquad (2.10)$$

Geometrically, the equivalent control can be constructed as

$$u_{eq} = \alpha u_+ + (1 - \alpha) u_- \qquad (2.11)$$

i.e. a convex combination of the values of $u$ on both sides of the surface $s(t)$. The value of $\alpha$ can be obtained from $\dot{s}(x) = 0$, requiring the system trajectories be tangent to the surface.

### 2.4.2 Lyapunov function approach

Based on choosing a candidate Lyapunov function:

$$V(x,t) = \frac{1}{2} \cdot s_i^T \cdot s_i \qquad (2.12)$$

Then, global reaching condition is given as:

$$\dot{V}(x,t) < 0 \qquad (2.13)$$

This reaching law results in a VSC where sliding mode is guaranteed only on the intersection of the switching surfaces. Finite reaching time is guaranteed if the above equation is modified as $\dot{V}(x,t) < -\epsilon$, $\epsilon$ is strictly positive.

### 2.4.3 The Direct switching function approach

The classic sufficient condition for sliding mode to appear is to satisfy the condition:

$$s_i . \dot{s}_i < 0, \quad i = 1, \ldots\ldots, m \tag{2.14}$$

Similarly, a substitute is proposed i.e.,

$$\lim_{s_i \to 0^+} \dot{s}_i < 0 \text{ and } \lim_{s_i \to 0^-} \dot{s}_i > 0 \tag{2.15}$$

These reaching laws result in a variable structure system where individual switching surfaces and their intersection are all sliding surfaces. This reaching is global but does not guarantee finite reaching time.

### 2.4.4 Gao's reaching law approach

Gao and Hung proposed a reaching law which directly specifies the dynamics of the switching surface by the differential equation, is represented in the generalized form as:

$$\dot{s} = -\varepsilon . sgn(s) - k . h(s) \tag{2.16}$$

where

$$\varepsilon = diag[\varepsilon_1, \ldots, \varepsilon_m], \quad \varepsilon_i > 0, i = 1,2, \ldots, m \tag{2.17}$$

$$k = diag[k_1, \ldots, k_m], \quad k_i > 0, i = 1,2, \ldots, m \tag{2.18}$$

and

$$sgn(s) = [sgn(s_1), sgn(s_2), \ldots, sgn(s_m)]^T \tag{2.19}$$

$$k(s) = [k(s_1), k(s_2), \ldots, k(s_m)]^T \tag{2.20}$$

$$s_i . k_i(s) > 0, k_i(0) = 0 \tag{2.21}$$

This theorem results in four special cases as:

#### 2.4.4.1 Constant Rate Reaching Law

$$\dot{s} = -\varepsilon . sgn(s), \varepsilon > 0 \tag{2.22}$$

where $\varepsilon$ represents a constant rate. This law constrains the switching variable to reach the switching manifold $s$ at a constant rate $\varepsilon$. Bounding the simplicity of the control law are the severe restrictions where the value of $\varepsilon$ if too small, reaching time will be too long whereas if $\varepsilon$ is too large, problem of chattering will dominate.

#### 2.4.4.2 Exponential Reaching Law

$$\dot{s} = -\varepsilon . sgn(s) - k . s, \varepsilon > 0, k > 0 \tag{2.23}$$

where, second term is the exponential term, having a solution as $s = s(0)e^{-kt}$. Therefore, by adding the proportional rate term $-k.s$, the state is forced to approach the switching manifolds faster when $s$ is larger.

### 2.4.4.3 Power Rate Reaching Law

$$\dot{s} = -k.|s|^\alpha.sgn(s), \ k > 0, 1 > \alpha > 0 \qquad (2.24)$$

This reaching law increases the reaching speed when the state is far away from the switching manifold. However, it reduces the rate when the state is near the manifold, resulting in a fast and low chattering reaching mode.

### 2.4.4.4 General Reaching Law

$$\dot{s} = -\varepsilon.sgn(s) - f(s), \ \varepsilon > 0 \qquad (2.25)$$

where, $f(0) = 0$, and $sf(s) > 0$ when $s \neq 0$.

After the selection of the reaching law, the control law can be determined using:

## 2.4.5 Augmenting the equivalent control

During sliding mode, equivalent control is derived using equivalent control method (Utkin, 1991). However, if the initial state of the system is not on the surface, then to force or drive the state trajectories onto the sliding surface $s$, discontinuous or switched control part is appended to the original equation, i.e.

$$u = u_{eq} + u_N. \qquad (2.26)$$

where, $u_N$ is added to satisfy the reaching condition which may have different forms.

## 2.4.6 Reaching Law method

By using any of the above mentioned four reaching laws, the control can be directly obtained by computing the time derivative $s(x)$ along the trajectory as:

$$\dot{s} = \frac{\partial s}{\partial x}.\dot{x} = \frac{\partial s}{\partial x}.(f(x) + g(x).u) = -k.s - \varepsilon.sgn(s) \qquad (2.27)$$

Resulting sliding mode is not reassigned but follows the natural trajectory on a first-reach-first-switch scheme. The switching takes place depending on the location of the initial state.

## 2.5 Chattering and its Reduction Techniques

To guarantee the desired behavior of the close-loop system, the sliding mode controllers require an infinitely fast switching mechanism. However, due to physical limitations in real world systems, directly applying the above developed control algorithms will always lead to high frequency oscillations in some vicinity of the switching surface, i.e., the so called chattering phenomenon [59-61]. There are two possible reasons behind chattering phenomenon. First, chattering may be caused by the switching nonidealities, such as time delays or time constants exist in microprocessor based switching devices. Second, even if the switching device is considered ideal and capable of switching at an infinite frequency, the presence of parasitic dynamics, i.e., unmodelled dynamics usually neglected in the open-loop model, also causes chattering to appear in the neighborhood of the sliding surface. However, in sliding mode controlled systems, due to the discontinuity of the control signal, the interactions between the parasitic dynamics and the switching term may result in a non-decaying oscillation with finite amplitude and frequency, i.e. chattering. If the switching gain is large, such kind of chattering may even cause unpredictable instability. Chattering and high control activity is the major drawbacks of the sliding mode approach in the practical realization of sliding mode control schemes. The chattering problem is considered as a major obstacle for SMC to become a more appreciated control method among practicing control engineers. In general, chattering must be eliminated for the controller to perform properly. Reducing this chattering effect has long been a major objective in research on SMC [64]. The existing approaches for chattering reduction in design of SMC are summarized below.

### 2.5.1 Boundary Layer method

A boundary layer around the sliding surface is specified. Inside the thin boundary layer neighboring the switching surface, the switching function is usually replaced by a linear feedback gain, thus the control signal becomes continuous and chattering is avoided.

$$B(t) = \{x, |s(x;t)| \leq \varphi \quad \varphi > 0 \quad (2.28)$$

where, $\varphi$ is the boundary layer thickness.

In other words, outside of $B(t)$, control law $u$ is chosen as before, which guarantees that the boundary layer is attractive, hence invariant. All trajectories starting inside $B(t = 0)$ remains inside $B(t)$ for all $T > 0$, allowing to interpolate $u$ inside the boundary layer) i.e. replacing the term $sign(s)$ $by$ $s/\varphi$ inside $B(t)$. The shortcoming of this approach is that the robustness properties of the sliding mode are actually lost inside the boundary layer, such that uncertainties and parasitic dynamics must be carefully considered and modeled in the feedback design in order to avoid instability. It also leads to tracking to within a guaranteed precision $\varepsilon$ (rather than perfect tracking). Saturation function 'sat(s)' is adopted instead of sign(s).

$$sat(s) = \begin{cases} 1 & s > \Delta \\ ks & |s| \leq \Delta, k = 1/\Delta \\ -1 & s < -\Delta \end{cases} \quad (2.29)$$

### 2.5.2 High-order sliding mode control

The control action is in this case a function of higher order time derivatives of the sliding variable. Second order sliding mode approach, used in this thesis, allows the definition of a discontinuous control $u$ steering both the sliding variable $s$ and its time derivative $\dot{s}$ to zero, so that the plant input $u$ is a continuous control and thus chattering can be avoided. Problem is that there is no general method for tuning the parameters which characterize the various algorithms.

### 2.5.3 Observer-based sliding mode control

This approach utilizes asymptotic state observers to construct a high frequency by pass loop, i.e., the control is discontinuous only with respect to the observer variables, thus chattering is localized inside the observer loop which bypasses the plant. This approach assumes that an asymptotic observer can indeed be designed such that the observation error converges to zero asymptotically.

### 2.5.4 VSC control with sliding sector

A Lyapunov function is used as an effective method to design a robust controller for uncertain systems. A Lyapunov function candidate is usually chosen as the square of the p-norm, i.e,

$$V = ||x||_p^2 = x^T.P.x > 0, \ x \neq 0 \quad (2.30)$$

where P is a positive definite symmetric matrix. It has been proved by the authors that for any controllable system, there always exists a special subset around a hyper plane, inside which the p-norm decreases, i.e., $\dot{V} \leq -x^T.R.x$ without needing any control action, where R is a positive semi-definite symmetric matrix. Such a subset is named as the PR-sliding sector [63]. One can use this property to design a Variable Structure controller such that outside the sliding sector, the VS control law is used to move the state into the sliding sector, and once the state is inside the sector, the Lyapunov function decreases with a specified velocity and zero input.

## 2.6 Modified Particle Swarm Optimization:

Particle swarm optimization (PSO) is the evolutionary computation technique proposed by Kennedy and Eberhart in 1995 [21]. Evolutionary algorithms are generic population based search approach, inspired by biological evolution i.e. reproduction, mutation, recombination and selection. Evolutionary Algorithms (EAs) are more elegant than traditional optimization techniques as they generate a population of potential solution rather than a single point solution. EAs have to proceed through several iterations and after every iteration, solutions having better fitness value are selected over poor solution and offspring are generated under some rule using these selected solutions.

### 2.6.1 Standard Particle Swarm Optimization (PSO):

Particle swarm optimization (PSO), derived from the social-psychological theory, is a type of evolutionary algorithm where each potential solution called 'particle' changes its position within the solution space with time, where each particle is the one-dimensional search space without the weight and size of the particle. During the iterations, each particle adjusts its position according to the best position obtained by the whole group as well as by that particular particle itself, termed as 'Velocity location search model'. The swarm direction of a particle is decided by the set of neighboring particle and the history of its own experience. Let $X_i$ and $V_i$ represent the $i$th particle position and its corresponding velocity in search space. The best previous position of $i$th particles recorded and represented as $Pb_i$. The best particle among all the particles in the group is represented as $Gb$. The updated velocity of $i$th individual is given as

$$V_i^{k+1} = w^k V_i^k + c_1 r_1 (Pb_i^k - X_i^k) + c_2 r_2 (Gb^k - X_i^k) \tag{2.31}$$

and the next generation value as follows

$$X_i^{k+1} = X_i^k + V_i^{k+1} \tag{2.32}$$

where, $V_i^k$- Velocity of $i$th individual at iteration $k$

$w^k$- Inertia weight at iteration $k$

$c_1 c_2$- Acceleration factors

$r_1 r_2$- Uniform random numbers between 0 and 1

$X_i^k$- Position of $i$th individual at iteration $k$

$Pb_i^k$- Best position of $i^{th}$ individual at iteration $k$

$Gb^k$- Best position of the group until iteration $k$

## 2.6.2 Modified Particle Swarm Optimization (MPSO):

To improve the convergence characteristics of the PSO algorithm, modifications have been implemented in it. Evolution equation of Modified PSO, proposed by Chang et. al [4], has been modified to ensure that the proposed algorithm has strong global convergence capability initially while it has strong local convergence capability later. In the Modified PSO algorithm $C_1$ decreases exponentially and $C_2$ increases exponentially with time. Effect of taking time dependent values for $C_1$ and $C_2$:

1. The dependence of particle's next position on its previous best position decrease with time.
2. The dependence of particle's next position on global best position increases with time.
3. The main strength of PSO is its property of having random particles but after sometime when the swarm has met some potential solutions and one of them is best possible till then. After this time, the swarm should search around that best solution only, because it is highly likely that the optimum value for fitness function is lying around near to that best solution.
4. It is highly unlikely that particles scattered around the full solution space will give better results than particles concentrated around a smaller subspace which contains the solution.
5. The whole swarm concentrates around the single global best point in very little time and continues their search for global optimum in a very small subspace around global best. Swarm concentrating around a small subspace results in better and quicker optimization.

The expressions for updating the velocity and position of each particle are

$$V_i^{(i+1)} = w * V_{in}^{(i)} + C_1 * \left(P_{in} - X_{in}^{(i)}\right) + C_2 * \left(G_{in} - X_{in}^{(i)}\right) \tag{2.33}$$

$$X_{in}^{(i+1)} = X_{in}^{(i)} + V_{in}^{(i+1)} \tag{2.34}$$

where,

$$w = 2 - \left(1 + \frac{1}{2k_{max}}\right)^i, \quad C_1 = e^{(-0.05t)} \text{ and } \quad C_2 = \frac{e^{(0.05t)}}{(1 + 0.05e^{(0.05t)})} \tag{2.35}$$

$i$ = current number particle,

$k_{max}$ = maximum number of iteration.

The objective or fitness function used here is integral squared error, as follows.

$$J(t) = \int_0^T e^2(t)dt \tag{2.36}$$

MPSO algorithm is used here to choose the parameters of PID-SMC method i.e. the switching and function parameters, which will be working off-line. It is generating a number of solutions for equal number of particles, and its quality will be evaluated by objective function. The objective function will take into account both the speed of reaching manifold and amplitude of chattering.

## 2.7 Modeling of Dynamic Systems

To achieve global asymptotic stability it was necessary to include some form of controller to deliver the system to the domain of attraction about which the linearized state space controller would achieve stability. To achieve global asymptotic stability it was necessary to include some form of controller to deliver the system to the domain of attraction about which the linearized state space controller would achieve stability.

### 2.7.1. INVERTED PENDULUM SYSTEM

The single inverted pendulum is a classical control problem in the field of non-linear control theory. The system shown in Figure 1 consists of an inverted pendulum mounted to a motorized cart. Applying a force to the cart, through a built in electrical motor, thus moving it forward and backward will cause the pendulum to swing. However, modeling these dynamics yields a highly non-linear problem. The main aim is to swing-up and stabilize the inverted pendulum upwards to the vertical position desired for the experiment and to maintain the position without letting it fall or change to an unstable state. One approach in finding a suitable control law is to linearize the system equation e.g. around the upright equilibrium position. As the linearized model does not represent the real model for large deviations, this approach is not suitable for a complete upswing and stabilization control. It is however required to design a nonlinear control design which performs the upswing movement and stabilizes it to the upright position. Sliding mode control is a very powerful nonlinear control design method and its advantages can be benchmarked using the example of a single inverted pendulum system. The inverted pendulum system considered here can be described by two nonlinear differential equations, derived using the Euler-Lagrange equation. Mathematical modeling for Figure 1 has been developed using free bode diagram, to obtain the dynamic motion equation. As seen from the Figure 1, there is no vertical motion so the equations of dynamic motion can be obtained by equating the sum of forces acting along the horizontal direction. The cart is driven by the dc motor, controlled by a controller. A disturbance force is applied on the top of the pendulum. To implement the sliding mode control algorithms, understanding the dynamic model of inverted pendulum is necessary and is developed after.

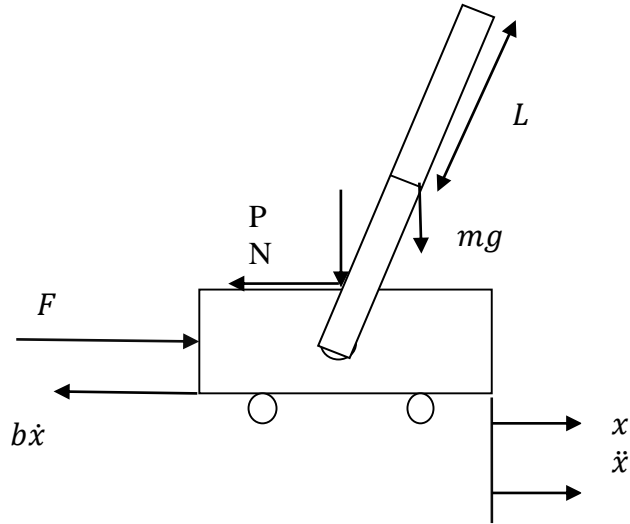

**Figure 2.1**. Inverted Pendulum with free body diagram

$$M\ddot{x} + b\dot{x} + N = F \tag{2.37}$$

$$\mathcal{T} = r\mathcal{F} = \mathbb{I}\ddot{\theta} \tag{2.38}$$

Horizontal component of the force = $ml\ddot{\theta}\cos(\theta)$. (2.39)

Component along the direction of N= $ml\dot{\theta}^2\sin(\theta)$ (2.40)

Horizontal component is given as

$$N = m\ddot{x} + ml\ddot{\theta}\cos(\theta) - ml\dot{\theta}^2\sin(\theta) \tag{2.41}$$

Substituting (2.41) in (2.37)

$$(M+m)\ddot{x} + b\dot{x} + ml\ddot{\theta}\cos(\theta) - ml\dot{\theta}^2\sin(\theta) = u \tag{2.42}$$

Vertical component on substitution is given as

$$P\sin(\theta) + N\cos(\theta) - mg\sin(\theta) = ml\ddot{\theta} + m\ddot{x}\cos(\theta) \tag{2.43}$$

Substituting (2.43) in (2.37)

$$(I + ml^2).\theta + mgl\sin(\theta) = -ml\ddot{x}\cos(\theta). \tag{2.44}$$

Therefore, the dynamic equation for the single stage inverted pendulum with considering b=0

$$\ddot{\theta} = \frac{mgl\sin(\theta) - m^2l^2\cos(\theta)\sin(\theta)\dot{\theta}^2 + u.ml\cos(\theta)}{m^2l^2\cos^2(\theta) - (I+ml^2)} \tag{2.45}$$

To build the nonlinear model of the Inverted Pendulum system for simulation purposes, physical modeling blocks of the Simscape extension 'SimMechanics' to Simulink is used. The blocks in the Simscape library represent actual physical components; therefore, complex multibody dynamic models can be built without the need to build mathematical equations from physical principles.

To achieve global asymptotic stability it was necessary to include some form of controller to deliver the system to the domain of attraction about which the controller would achieve stability. Sliding mode control is quite capable of controlling the pendulum over all angles and is therefore suitable for swing-up and stabilization control.

**Table 2.1**
**Operating Parameters of the Inverted Pendulum system**

| S. No. | Parameters | Description of the Quantity | Value with units |
|---|---|---|---|
| 1 | Mc | Mass of the cart | 1 kg |
| 2 | Mp | Mass of the pendulum | 0.1 kg |
| 3 | I | Moment of Inertia | 0.006 kgm$^2$ |
| 4 | L | Length of the pendulum | 0.3 m |
| 5 | G | Acceleration due to gravity | 9.8 ms$^{-2}$ |
| 6 | B | Frictional coefficient | 0 |

The equation relating force '$F$' applied to the cart and the motor voltage '$V_m$' applied to the DC motor is defined as

$$F = \alpha(V_m - K_m(K_g \cdot \dot{x}/r_m)) \tag{2.46}$$

where $K_m$, $K_g$, $r_m$ are respectively the motor constants, the gear reduction ratio, the radius of the spur gear that moves the cart via the motor and a constant $\alpha = \eta_g K_g \eta_m K_t / R_m r_m$. The terms $\eta_g$, $\eta_m$, and $R_m$ are respectively the gear reducer efficiency, the motor efficiency, and the motor resistance.

### 2.7.2. NONLINEAR EXAMPLE

Van der pol's equation [75] is used as a test problem in the last example, for simulation and to verify the proposed algorithm. Van-der-Pol (VDPO) oscillator is used for representing non-linear deterministic systems in delta domain. VDPO oscillator is an oscillator with non-linear damping and it evolves for the control of the position. The dynamics of non-linear system is

$$\dot{x}_1 = x_2$$

$$\dot{x}_2 = -2x_1 + 3(1 - x_1^2)x_2 + u \qquad (2.47)$$

$$y = x_1$$

## 2.8 Conclusion

Aim of the controller

- Design a control law to effectively account for: the parametric uncertainty (imprecision on mass properties, inaccuracies on torque constant), and the presence of unmodelled dynamics (neglected time-delays etc.).
- Quantify the resulting modeling/performance trade-offs, and in particular, the effect on tracking performance of discarding any particular term in the dynamic model.

Sliding mode control methodology is based on a notational simplification, which amounts to replacing an $n^{th}$ order tracking problem by a $1^{st}$ order stabilization problem. The main advantages of the sliding mode control approach are the simplicity of design and implementation, the high efficiency and the robustness with respect to matched uncertainties. However, it has been shown that imperfections in switching devices and delays were inducing a high frequency motion called chattering. Chattering and high control activity was the reasons that fomented a generalized criticism towards sliding mode control. To avoid chattering some approaches were proposed. The main idea was to change the dynamics in a small vicinity of the discontinuity surface in order to avoid real discontinuity and at the same time preserve the main property of the whole system. However, the trajectories of the controlled system remain in a small neighborhood of the surface and the robustness of the sliding mode was partially lost. The most recent and interesting approach for the elimination of chattering is represented by the second order sliding mode methodology.

# Chapter 3

# Design of PID-type first order Sliding Mode Controller for Non-linear Systems

## 3.1 Introduction

Problem of Variable structure system (VSS) analysis lead to the solution of a differential equation with discontinuous right hand side. Variable structure control results from multiple control structure. The state feedback control law is discontinuous in nature as it switches from one continuous structure to another respective to the evolution of the state vector, confining the system dynamics according to the applied feedback controller as well as to the switching strategy. This multiple control structure acts like a parallel connection of several different continuous subsystems, acting one at a time. By considering the nonlinear dynamic system

$$\dot{x} = f(x,t,u) + g(x,t).u(t) \qquad x \in \mathcal{R}^n, \; u \in \mathcal{R}^m. \tag{3.1}$$

subjected to discontinuous feedback

$$u_i = \begin{cases} u_i^+(t,x) \; if & s_i(x) > 0 \\ u_i^-(t,x) \; if & s_i(x) < 0 \end{cases} \quad \text{if} \quad i=1,2,\ldots,m. \tag{3.2}$$

Sliding mode control is a fundamental property of Variable structure system (VSS) as it exhibit a particular behavior characterized by the commutation between various system structures taking place at infinite frequency. Sliding mode control alters the dynamics of a nonlinear system by applying a multiple feedback control structure acting on the opposite sides of a predetermined surface in the system state space. Each of the controllers is designed so as to ensure that the representative trajectory is pushed towards the surface, so that it approaches the surface and once it hits the surface for the first time, it stays on it afterwards. The resulting motion of the system is restricted to the surface which is interpreted as 'sliding' of the system along the surface. This motion of the system as it slides along the manifold is called sliding mode and the geometrical locus is called sliding surface. Sliding surface is defined as

$$S(x,t) = (\frac{d}{dt} + \lambda)^{(n-1)}.e \qquad (3.3)$$

Sliding surface is described by $s = 0$ and the sliding mode commences after a finite time when the system trajectories reached the surface. Sliding mode is characterized by a very high robustness feature and is insensitive to model uncertainties and disturbances. Single input nonlinear systems second order equation is represented as

$$\ddot{\theta} = f(\theta, \dot{\theta}, t) + g(\theta, t).u(t) + d(t) \qquad (3.4)$$

where, $f(\theta, \dot{\theta}, t)$ and $g(\dot{\theta}, t)$ are known nonlinear functions, embodying plant parameters, model uncertainties and unmodelled dynamics.

$d(t)$ = bounded external disturbances.

t = independent time variable.

$\theta$ = output at any particular time, $\theta(t) \epsilon \mathcal{R}$.

$u(t)$ = control action generated.

The control problem is to asymptotically track the desired response using a discontinuous control law generated by the sliding mode algorithm, in the presence of parameter variations, model uncertainties and bounded external disturbances. To track the desired trajectory, error between the desired and the reference trajectory is:

$$e(t) = \theta_r(t) - \theta(t) \qquad (3.5)$$

In the first order sliding mode controller design, sliding surface is chosen with a relative degree of one w.r.t. the control input, i.e. control input acts on the first derivative of the sliding surface. Higher order sliding modes is the most popular method of chattering attenuation and control input acts on the higher order derivatives of the sliding surface. It also provides smooth control and better performance in its implementation yielding less chattering and faster convergence time while retaining the robustness properties. Convergence entails the problem of stability of the origin in an m-dimensional manifold, therefore existence condition has to be formulated in terms of generalized stability theory, hence the second method of Lyapunov used for analysis. This requires selecting a generalized Lyapunov function $V(t, x)$ which should be positive definite and should have negative time derivative in the region of attraction [66-69].

## 3.2 PID-type sliding surface based Sliding mode Control

To generate a control for the defined PID sliding surface, which has Proportional, Integral and derivative error term, will remove the deficiency of traditional Sliding Mode Control. This PID sliding surface is defined as:

$$s(t) = K_p e(t) + K_d \dot{e}(t) + K_i \int_o^t e(\mathcal{T}) d\mathcal{T} \qquad (3.6)$$

where $K_p, K_i, K_d$ are positive coefficients providing flexibility for sliding surface design and $K_p, K_i, K_d \in \mathcal{R}^+$ and $s(t) \in \mathcal{R}$. Assuming that the system is initiated at the region $s(t) > 0$, will result in increasing $s(t)$ and an insufficient input to drive the error to converge to the sliding manifold. Integral action increases the control action to force the error to the siding manifold leveling with the reaching condition $\dot{V}(t) < 0$. Control action is reduced as $s(t) \to 0$ forcing the system into the sliding mode. This problem of forcing to the sliding surface, constraining the error towards zero is equivalent to that of set point tracking for $t > 0$. Proportional and derivative action acts on the stabilization part of the control action. The control forces for the early convergence of the error to the sliding manifold, which in turn amplifies the integral step leading to a better switching manifold convergence.

A first order sliding mode control is defined if and only if $s(t) = 0$ and fundamental criterion as $s(t).\dot{s}(t) < 0$. Aim of the controller, is to force the tracking error $e(t)$ to converge on the sliding surface $s(t) = 0$. By maintaining the error e(t) onto the sliding surface $s(t)$ will lead to e(t) approaching zero as $t \to \infty$. After reaching the sliding surface s(t), error will simply slide along the sliding hyper-surface, into the origin e(t)=0 even in the presence of matched parameter uncertainties, model uncertainties and bounded disturbances. The error after starting with an initial value will converge to the boundary layer and slide along it until it reaches $\dot{e}(t) = 0$.

First order derivative of the sliding manifold (PID surface) is represented as

$$\dot{s}(t) = K_i e(t) + K_p \dot{e}(t) + K_d \ddot{e}(t) \qquad (3.7)$$

On deriving the error to second order, we get

$$\dot{e}(t) = \dot{\theta}_r(t) - \dot{\theta}(t) \tag{3.8}$$

$$\ddot{e}(t) = \ddot{\theta}_r(t) - \ddot{\theta}(t) \tag{3.9}$$

Substituting the error equation of $\ddot{\theta}(t)$ into the second derivative and writing the resultant first derivative surface equation, we get

$$\dot{s}(t) = K_i(\theta_r(t) - \theta(t)) + K_p(\dot{\theta}_r(t) - \dot{\theta}(t)) + K_d(\ddot{\theta}_r(t) - \ddot{\theta}(t)) \tag{3.10}$$

Control input is defined as

$$u(t) = u_{eq} + u_{sw} \tag{3.11}$$

where, $u_{eq}$ = Equivalent / Reaching control, and

$u_{sw}$ = Switched control.

### 3.2.1 Equivalent control

This theory proposed by Utkin [44] is based on system with estimated plant parameters with zero disturbances. It is defined as the smooth control law that can be used to determine the system motion restricted to the switching surface $s(x) = 0$, when initial error is precisely located on the sliding surface $s(t)$. For equivalent control is written as $\dot{s}(t) = 0$ with $d(t, u(t)) = 0$.

$$\dot{s}(t) = K_i(\theta_r(t) - \theta(t)) + K_p(\dot{\theta}_r(t) - \dot{\theta}(t)) + K_d(\ddot{\theta}_r(t) - \ddot{\theta}(t)) = 0 \tag{3.12}$$

On substituting (3.4) into (3.12) result in,

$$\dot{s}(t) = K_i(\theta_r(t) - \theta(t)) + K_p(\dot{\theta}_r(t) - \dot{\theta}(t)) + K_d(\ddot{\theta}_r(t) - f(\theta, \dot{\theta}, t)) - g(\theta, t).u(t) = 0 \tag{3.13}$$

Therefore, Equivalent control is given as

$$u_{eq} = -\frac{1}{K_d g(\theta, t)}\left(K_i(\theta_r(t) - \theta(t)) + K_p(\dot{\theta}_r(t) - \dot{\theta}(t)) + K_d(\ddot{\theta}_r(t) - f(\theta, \dot{\theta}, t))\right) \tag{3.14}$$

where, $\theta(t)$ =constant according to the requirement.

$K_p, K_i, K_d$ have to selected properly so that the tracking error converge to zero, constraining to the sliding surface leading to a condition where the polynomial $(K_i e(t) + K_p \dot{e}(t) + K_d \ddot{e}(t) = 0)$ should be Hurwitz.

Hurwitz polynomial defines that the roots must lie on the left side of the plane implying $\lim_{t \to \infty} e(t) = 0$ means globally asymptotic stability for the closed loop system:

$e(t) \to 0$ as $t \to \infty$.

## 3.2.2 Exponential Power rate reaching law design

Sliding mode control law is based on the Reaching law comprising the reaching phase that drives the system to a stable manifold and the sliding phase which ensures system to slide to the equilibrium. Constant rate reaching law and Exponential reaching laws are defined as

$$\dot{s} = -k_1 s - \varepsilon_1 sign(s) \tag{3.15}$$

To improve the convergence capability of the equivalent control approach, reaching laws [80] are used defined in [21-26]. The Proposed scheme is termed as 'Power rate exponential reaching law', is defined as:

$$\dot{s} = -ks - k_{sc}|s|^\alpha . sat(s) \tag{3.16}$$

where $k > 0, k_{sc} > 0, 0 \leq \alpha \leq 2$ and sat(s) from (27).

By satisfying the reaching condition (3.16), the phase trajectory of the system can reach the sliding manifold in finite time and stays at that state. In the reaching condition, the reaching speed will become faster as the term become larger, but will also results in more chattering in the neighborhood of the sliding manifold $s = 0$. By decreasing the magnitude of the reaching condition term, finite reaching time will increases due to the reduction in the convergence velocity leading to lower chattering in the neighborhood of the sliding manifold $s = 0$. To prove its robustness characteristics, combine (3.16) and (3.12) will result in

$$\dot{s}(t) = K_i\big(\theta_r(t) - \theta(t)\big) + K_p\left(\dot{\theta}_r(t) - \dot{\theta}(t)\right) + K_d\left(\ddot{\theta}_r(t) - \ddot{\theta}(t)\right) = -ks - k_{sc}|s|^\alpha . sat(s) \tag{3.17}$$

Substituting (3.16) in (3.13) results in

$$K_i\big(\theta_r(t) - \theta(t)\big) + K_p\left(\dot{\theta}_r(t) - \dot{\theta}(t)\right) + K_d\left(\ddot{\theta}_r(t) - f(\theta,\dot{\theta}, t) - g(\theta,t).u(t)\right) = -ks - k_{sc}|s|^\alpha . sat(s) \tag{3.18}$$

Interpolating the equation results in the expression of the controller as

$$u(t) = -(K_d . g(\theta,t))^{-1} . (K_i\big(\theta_r(t) - \theta(t)\big) + K_p\left(\dot{\theta}_r(t) - \dot{\theta}(t)\right) + K_d\left(\ddot{\theta}_r(t) - f(\theta,\dot{\theta}, t)\right) + ks + k_{sc}|s|^\alpha . sat(s)) \tag{3.19}$$

## 3.3 Stability

**3.3.1** Lyapunov stability criteria's origin dates back to 1970's by Leitmannn and Gutman [38, 67-68]. It is the most popular approach to evaluate and to prove the stable convergence property of the sliding mode controller. Second Lyapunov approach is used to prove the stability. Candidate lyapunov function is given as:

$$V(t) = \frac{1}{2}s^2(t) \text{ with V(0)=0 and V(t)>0 for s(t)} \neq 0. \tag{3.20}$$

A sufficient condition or reaching condition is to select the control law that forces the trajectory of error to the sliding phase from reaching phase, given as

$$\dot{V}(s) = -s^T diag\{sign(s_i)\}|\dot{s}| = -|s^T||\dot{s}| < 0 \tag{3.21}$$

Substituting (16) in (21), we get

$$\dot{V}(t) = s(t)(-k_{sc}sign(s(t)) - d(t)) \tag{3.22}$$

which is derived to prove the Lyapunov criteria.

$$\dot{V}(t) = -k_{sc}|s(t)| - s(t)d(t) \tag{3.23}$$

$$\dot{V}(t) \leq -k_{sc}|s(t)| + s(t)d_{max} \tag{3.24}$$

$$\dot{V}(t) \leq -|s(t)|(k_{sc} - d_{max}) \tag{3.25}$$

$$\dot{V}(t) \leq 0. \tag{3.26}$$

In the reaching phase with $s(t) \neq 0$, we get $k_{sc} > d_{max}$ and $|s(t)| > 0$ which lead to $\dot{V}(t) \leq 0$ i.e. a negative definite condition, satisfying the direct Lyapunov stability criteria.

**3.3.2.** Candidate Lyapunov function

$$V(t) = {}^1/_2 s^2 => \dot{V}(t) = s\dot{s} \tag{3.27}$$

On substituting,

$$\dot{V}(t) = s|-ks - k_{sc}|s|^\alpha . sat(s)| \tag{3.28}$$

$$\dot{V}(t) = -ks^2 - k_{sc}|s|^\beta. \tag{3.29}$$

So $\dot{V}(t) < 0$ because $> 0, k_{sc} > 0$, verifying the condition that the PID sliding mode surface exist and can reach under the control law.

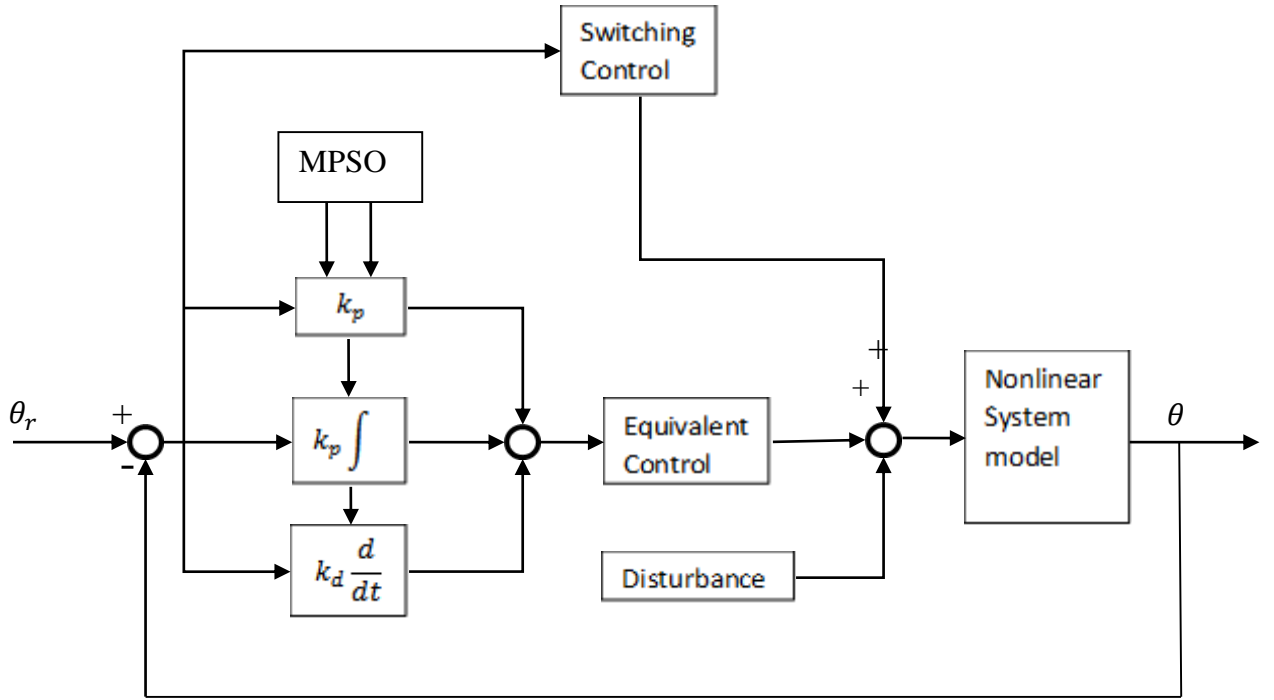

**Figure 3.1** Description of the nonlinear system based on sliding mode control.

## 3.4 Result and Discussion

The Modified PSO algorithm is applied to find the parameters of the above system when the system is initially at stable equilibrium position i.e. vertically downward position. These parameters are obtained offline for the proposed PID sliding surface based sliding mode controller using improved exponential power rate reaching law (50, 60) shown in Figure 3.1. In the simulations, the variables used in the modified PSO algorithm are given by Parameters of sliding modes, which were held constant during experiments. PID Sliding mode control is of general form given by $u(t)$ Control force, which is defined in terms of the error between the desired and actual output and $K_d, K_i, K_p, k\ and\ k_{sc}$ are switching parameters. PID and switching parameters are chosen such that the proportional gain $K_p$ has positive values to improve the activity, and the derivative gain $K_d$ limited to suppress the high frequency and noise, and value of $k$ should be greater than $k_{sc}$ to reduce the effect of chattering Setting used in Matlab implementation of MPSO algorithm are: Number of Particles, n=50, Number of subpopulation=5, Maximal number of iterations=90 and the search interval for each particle is given by [-5 5].

*Example 1: Inverted Pendulum system*

Dynamic equation for the single stage inverted pendulum, considering b=0 is

$$\ddot{\theta} = \frac{mgl\sin(\theta) - m^2l^2\cos(\theta)\sin(\theta)\dot{\theta}^2 + u.ml\cos(\theta)}{m^2l^2\cos^2(\theta) - (I+ml^2)} \quad (3.30)$$

Substituting the above model equation to the control equations, we get

Equivalent control equation:

$$u_{eq} = -\frac{1}{K_d g(\theta,t)}\left(K_i(\theta_r(t) - \theta(t)) + K_p(-\dot{\theta}(t)) + K_d(-f(\theta,\dot{\theta},t)) - sgn(s(x))\right) \quad (3.31)$$

Reaching law control equation:

$$u(t) == -(K_d \cdot g(\theta,t))^{-1} \cdot (K_i(\theta_r(t) - \theta(t)) + K_p(\dot{\theta}_r(t)) + K_d(\ddot{\theta}_r(t) - f(\theta,\dot{\theta},t)) + ks + k_{sc}|s|^\alpha \cdot sat(s)) \quad (3.32)$$

The Inverted Pendulum system having the parameters is given in Table 1 is considered for swing up and stabilization with disturbances acting on the system directly. Simulations are conducted on the SimMechanics model of the Inverted Pendulum and the results provide the effectiveness of the proposed algorithm. The MPSO PID sliding surface based sliding mode controller using improved reaching law is applied to the above system when the pendulum is initially at stable equilibrium point, i.e. at the vertical downward position making it the initial condition. The values of the parameters are chosen as $K_d = 0.8, K_i = 4, K_p = 105, k = 35$ and $k_{sc} = 1.5$, and the desired position is at vertically upwards position i.e. $\theta = 0$. External disturbance applied to the IP system is $d = 10\sin(t)$.

Simulations is performed for two cases:

Firstly, the Inverted Pendulum example is considered for the stabilization part where the system is initially at its vertically upward position, with the disturbance acting on the system given by $d = 1000\delta(10t)$.

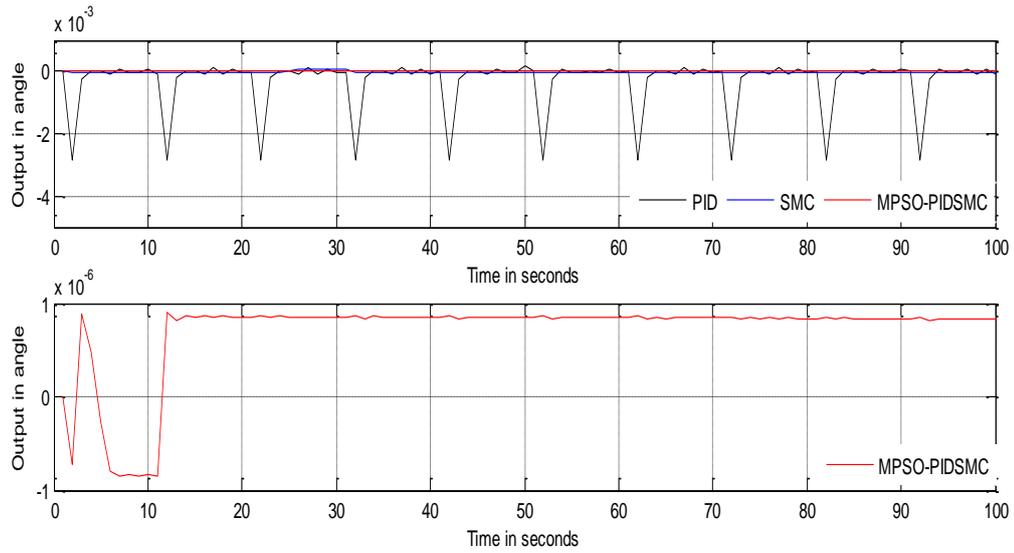

**Figure 3.2** Simulations result: Output comparison for PID (3), SMC (84, 52) and proposed MPSO PID-SMC with uncertainty as external disturbance, d= $1000\delta(10t)$, and Output of MPSO PID SMC with improved reaching law.

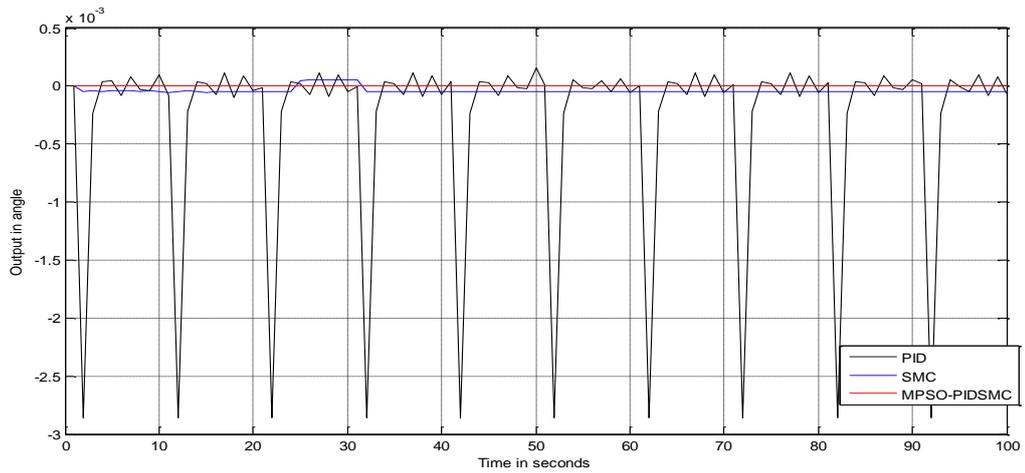

**Figure 3.3** Simulations result: Output comparison for PID (3), SMC (76, 50) and proposed MPSO PID-SMC with uncertainty as external disturbance, d= $1000\delta(10t)$.

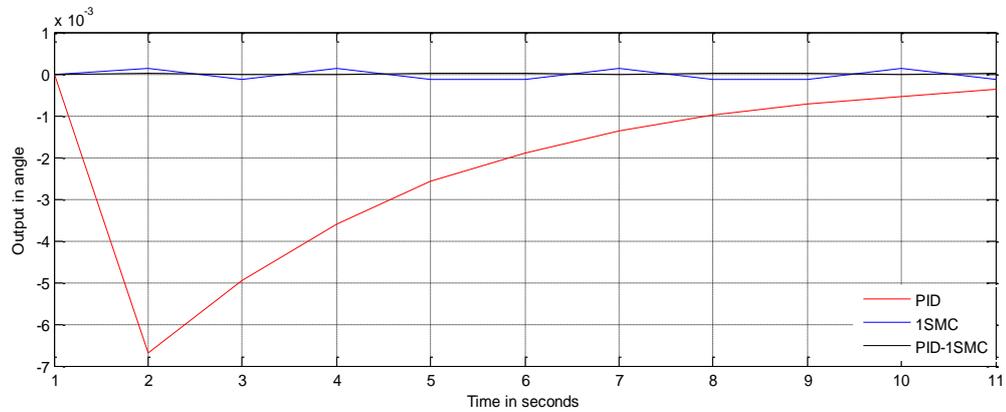

**Figure 3.4** Simulations result: Transient output comparison for PID (3), SMC (76, 50) and proposed MPSO PID-SMC with uncertainty as external disturbance, d= $1000\delta(10t)$.

Figure 3.2 shows the performance comparison between controllers PID [3], SMC [76, 52] and the proposed MPSO PID sliding surface based sliding mode control with improved reaching law, which clearly depicts invariance property of SMC w.r.t. PID and better convergence of the PID-SMC w.r.t. SMC.

Secondly, Inverted Pendulum system is considered is considered for the swing-up and stabilization where the system is initially at its downwards equilibrium position, with disturbances acting on is given by $d = 10\sin(t)$.

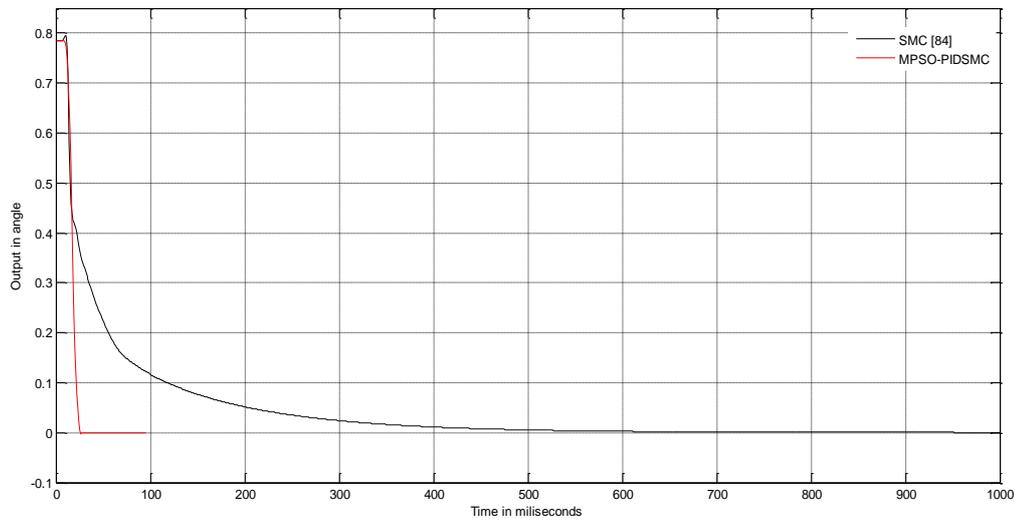

**Figure 3.5** Simulations result: Transient response comparison for SMC [76, 50] and proposed MPSO PID-SMC with improved reaching law and uncertainty as external disturbance, d= $10\sin(t)$.

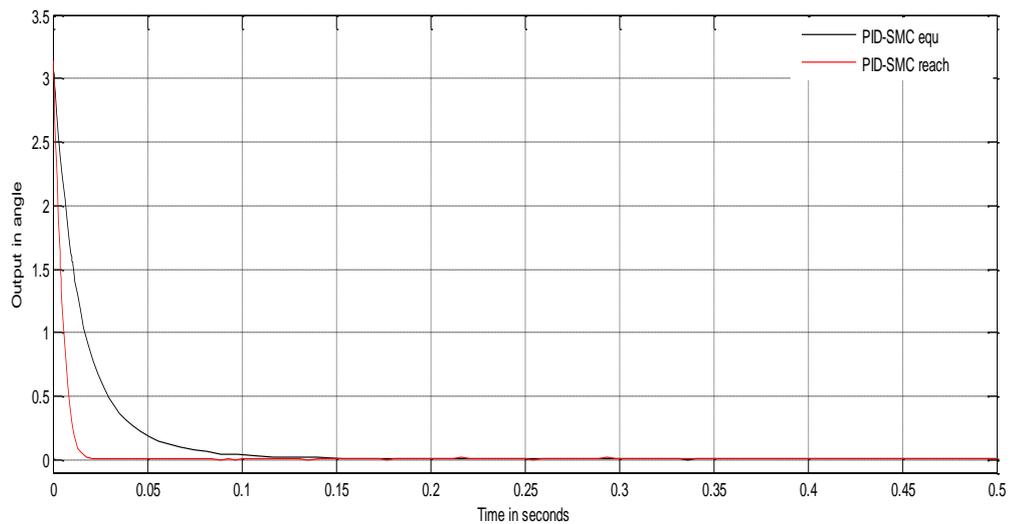

**Figure 3.6** Simulation result: Output with uncertainty 'd' for MPSO PID-SMC with improved reaching law compared with equivalent control method [39].

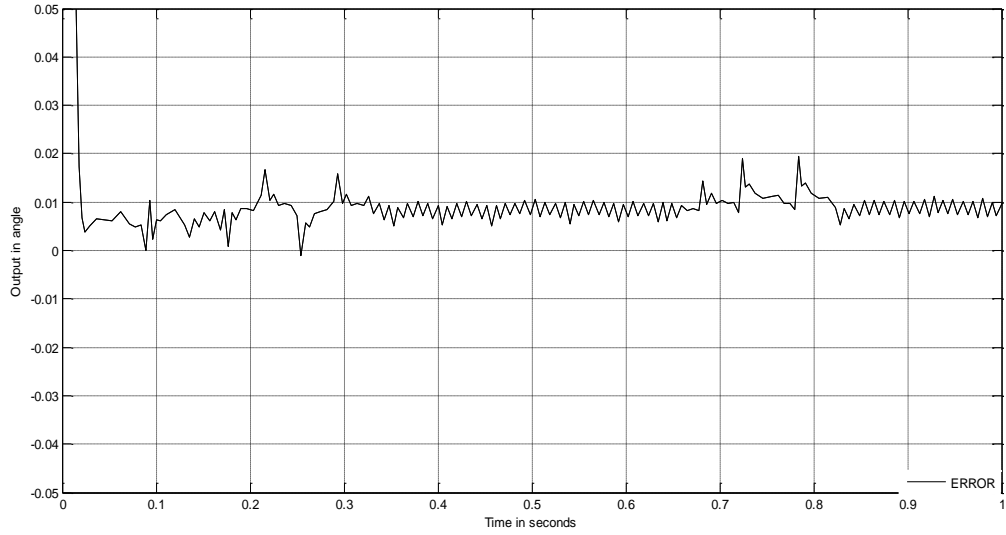

**Figure 3.7** Simulations result: Error with uncertainty 'd' for MPSO PID-SMC with improved reaching law.

In Figure 3.5 proposition of faster convergence to the error surface is shown by simulation which also proves the usefulness of the Integral term in the sliding surface, as described above. In Figure 3.5 the performance of the second order Inverted Pendulum is shown when an uncertainty defined by d= $10\sin(t)$ is applied. The sliding mode control uses the property of order reduction where the order of the system is reduces by one when the system lies on the sliding surface. Integrator part of the sliding surface forces the system to converge to the error surface at fast pace. Figure 3.7 shows that the steady state error is very near to the zero error surfaces limiting itself to the boundary upon which the sliding mode's chattering phenomenon is forced.

**Table 3.1**
**Comparison of Simulation results for the Inverted Pendulum system**

| S.No. | Parameters | 1-SMC | PID-SMC (eq.) | Proposed PID-SMC |
|---|---|---|---|---|
| 1 | Rise time (seconds) | 0.35 | 0.1 | 0.02 |
| 2 | Settling time (seconds) | 0.55 | 0.125 | 0.02 |

*Example 2: Van der Pol Equation*

Control effort required for tracking control of the desired level process is given as

$$u(t) = -(K_d \cdot g(\theta, \dot{\theta}, t))^{-1} \cdot (K_i(\theta_r(t) - \theta(t)) + K_p(\dot{\theta}_r(t) - \dot{\theta}(t)) + K_d(\ddot{\theta}_r(t) - f(\theta, \dot{\theta}, t)) + ks + k_{sc}|s|^\alpha \cdot sat(s)) \quad (3.33)$$

Considering the system (2.47) where the external disturbance $d = 10\sin(t)$. Following are the extensive simulations to demonstrate the effectiveness of the proposed PID sliding mode tracking control. To proceed the design of PID sliding mode control, parameters chosen as $K_d = 0.8, K_i = 8, K_p = 105, k = 35 \text{ and } k_{sc} = 1.5$. The initial conditions are chosen as [pi/60 0], and the desired trajectory is chosen as $y_d = 0.1\sin(t)$.

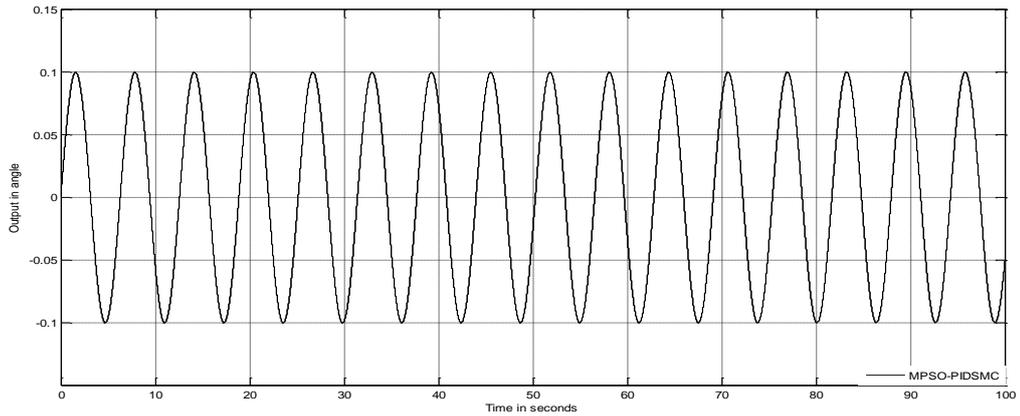

**Figure 3.8** Tracking control for the van der pol equation using derived PID switching parameters.

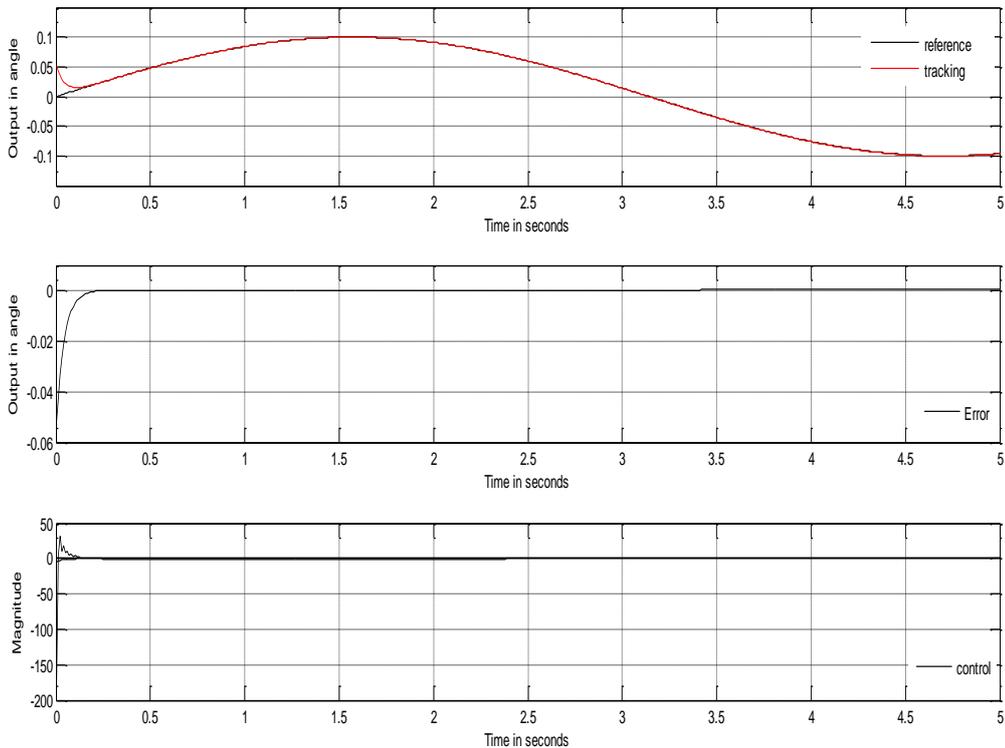

**Figure 3.9** Simulatios results: with disturbance 'd'. (a)Tracking control for the van der pol equation using the PID sliding mode control with $y_d = 0.1\sin(t)$. (b) error output. (c) control force.

In the following simulations, matlab\simulink 13 is utilized to implement the above MPSO PID-SMC based improved reaching law algorithm. The sampling time is set to 0.01 seconds for simulating the nonlinear differential equations. The comparison of MPSO PIDSMC with the SMC is shown in Figure 3.5. Figure 3.9(a) shows excellent tracking performance for the given system with a very less convergence time and complete tracking even in the presence of applied disturbace, shown in Figure 3.8. Figure 3.9(b) stamped the concept of finite convergence and tracking phenomenon. Figure 3.9(c) shows the required control effort applied to the system.

## 3.5 Conclusion

Modified Particle Swarm Optimization (MPSO) PID sliding surface based Sliding mode control with exponential power rate reaching law is proposed for controlling non-linear systems. The problem of swing up and stabilization of the Inverted Pendulum, which is a benchmark example for control of nonlinear systems, is considered for the application of the proposed algorithm. Simulations studies confirms the improved performance of the proposed PID-SMC ensuring invariance property in the presence of matched uncertainties and external disturbances, compared to the first order sliding mode and PID controllers, without sacrificing the tracking accuracy. The closed loop system has been shown to be stable in the sense of direct Lyapunov's approach. In the experimental example of conical tank, this algorithm has been implemented showing the effectiveness of the proposed control method. In order to avoid chattering phenomenon, saturation function (boundary condition) has been adopted to smoothen the switching signal. The PID-SMC guarantees better transient performance and in fact produces faster system response. The steady state error is smaller than the SMC found its suitability for tracking purposes, also better behavior of the output in case of bounded external disturbance. Experimental results confirm the fact that the sliding mode controllers are reasonable candidate to be used in industrial applications as they simple to use and easy to implement.

\

# Chapter 4

# Second Order Sliding Mode PI-PD Controller for Inverted Pendulum

## 4.1 Introduction

In this chapter, a Second Order Sliding Mode PI-PD Controller is designed for Inverted Pendulum which helps to avoid the chattering problem exists in first order sliding mode control. A PI (Proportional-Integral) type surface based second order sliding mode controller is designed using modified reaching law to improve the performance for nonlinear state differential equations with unknown parameters. To remove the problem of singularity and peak overshoot, a PD type sliding surface based sliding mode controller is connected through the feedback. This paper highlights on the feedback compensator design for different sliding surface based sliding mode controller and the important features of multiplexing different control inputs resulting in robustness and higher convergence of output. Stability study of the system has been done by using direct Lyapunov stability criterion.. Simulations and experimental application is done on the Inverted Pendulum system for the case of swing up and stabilization as well as tracking (with external disturbance), to evaluate the controller performance, complexity of implementation and stability analysis.

## 4.2 Controller Structure

A PI sliding surface is proposed for the second order sliding as,

$$s(t) = K_p e(t) + K_i \int_0^t e(\mathcal{T})d\mathcal{T} \tag{4.1}$$

where $K_p, K_i$ are positive coefficients providing flexibility for sliding surface design and $K_p, K_i \in \mathcal{R}^+$ and $s(t) \in \mathcal{R}$. Assumption being that system is initiated at the region $s(t) > 0$, resulting in increasing $s(t)$ and an insufficient input to drive the error to converge to the sliding manifold. Integral action increases the control action to force the error to the siding manifold. Control action is reduced as $s(t) \to 0$ forcing the system into the sliding mode. A sliding mode design is called as 2-sliding mode control if $\ddot{s}(t) = 0$. Second order sliding surface, written as:

$$\dot{s}(t) = K_p \dot{e}(t) + K_i e(t) \tag{4.2}$$

$$\ddot{s}(t) = K_p \ddot{e}(t) + K_i \dot{e}(t) \tag{4.3}$$

On substitution,

$$\ddot{s}(t) = K_p(\ddot{\theta}_r(t) - \ddot{\theta}(t)) + K_i \dot{e}(t) \tag{4.4}$$

$$K_p\left(\ddot{\theta}_r(t) - f(\theta,\dot{\theta}) - g(\theta).u(t)\right) + K_i \dot{e}(t) = -k_1 \dot{s} - k_2 s - \varepsilon_1 |s|^\alpha sign(s) - \varepsilon_2 |s|^\alpha sign|\dot{s}| \tag{4.5}$$

Therefore, the control input is

$$u(t) = \left(K_p g(\theta)\right)^{-1}\left(-K_p \ddot{\theta}_r(t) + K_p f(\theta,\dot{\theta})\right) - K_i \dot{e}(t) - k_1 \dot{s} - k_2 s - \varepsilon_1 |s|^\alpha sign(s) - \varepsilon_2 |s|^\alpha sign|\dot{s}|) \tag{4.6}$$

Resulting control input results in an improved response characteristics with fast convergence time. It also results in removal of chattering characteristics. But response due to PI sliding surface leads to a peak overshoot as well as the problem of singularity (Fig 3b). First order derivative of the sliding manifold (PD surface) is represented as $(K_d = 1)$

$$\dot{s}(t) = K_p \dot{e}(t) + \ddot{e}(t) \tag{4.7}$$

$$\dot{s}(t) = K_p\left(\dot{\theta}_r(t) - \dot{\theta}(t)\right) + (\ddot{\theta}_r(t) - \ddot{\theta}(t)) \tag{4.8}$$

$$K_p\left(\dot{\theta}_r(t) - \dot{\theta}(t)\right) + \left(\ddot{\theta}_r(t) - f(\theta,\dot{\theta}) - g(\theta).u(t)\right) = -k_1 s - \varepsilon_1 |s|^\alpha sign(s) \tag{4.9}$$

$$u(t) = -g(\theta)^{-1}(-\ddot{\theta}_r(t) + f(\theta,\dot{\theta})) - K_p \dot{e}(t) - k_1 s - \varepsilon_1 |s|^\alpha sign(s)) \tag{4.10}$$

To negate the flaws of the above proposed technique, a PD sliding surface is connected as negative feedback leading to the removal of the problem of singularity as well as markedly improves the response characteristics.

## 4.3 Modified Reaching Law

Sliding mode control law is based on the Reaching law comprising the reaching phase that drives the system to a stable manifold and the sliding phase which ensures system to slide to the equilibrium. Constant rate reaching law and Exponential reaching laws are defined [11] as

$$\dot{s} = -k_1 s - \varepsilon_1 sign(s) \tag{4.11}$$

To improve the performance of the control law behavior, the reaching law for the second order sliding mode is modified as:

$$\ddot{s} = -k_1 \dot{s} - k_2 s - \varepsilon_1 |s|^\alpha sign(s) - \varepsilon_2 |s|^\alpha sign|\dot{s}| \tag{4.12}$$

where, $k_i > 0, \varepsilon_i > 0, 0 \leq \alpha \leq 2$.

In the above equations, $\varepsilon_i$ term only forces the variable to reach the sliding manifold $s$ at a constant rate. Value of $\varepsilon_i$ has to take smaller to eliminate the effect of chattering, but will lead to a less convergence speed. The proportional term forces the state to approach the switching manifold faster for larger $s$. But a larger term may result in chattering and a larger control value. Power term increases the reaching speed when the state is far away and reduces when the state is near the switching manifold. The power rate sign term generates the control force which guarantees the appearance of a 2-sliding mode attracting the trajectories of the sliding variable dynamics in finite time. For the second order surface, reaching law is modified to make the control continuous as well as to attenuate chattering. Similarly, the first order PD sliding mode reaching law is modified as,

$$\dot{s} = -k_1 s - \varepsilon_1 |s|^\alpha sign(s) \tag{4.13}$$

## 4.4 Stability Analysis

The instability in the system disturbs the performance. So, stabilization is required to improve the performance of the system. Lyapunov stability approach is used to prove the stable convergence of the nonlinear system. Most popular among is direct Lyapunov stability approach to investigate the stability property of the proposed second order sliding mode algorithm.

Considering a Lyapunov candidate function as:

$$V = \tfrac{1}{2} s^2 + \tfrac{1}{2} \dot{s}^2 \qquad s(t) \neq 0, \dot{s}(t) \neq 0 \tag{4.14}$$

On differentiating, to satisfy the direct Lyapunov function condition

$$\dot{V} = s\dot{s} + \dot{s}\ddot{s} \tag{4.15}$$

To guarantee the stability of the system, derivative of the Lyapunov function should be negative definite, i.e. $\dot{V} \leq 0$.

$$\dot{V} = s\dot{s} + \dot{s}(-k_1\dot{s} - k_2 s - \varepsilon_1|s|^\alpha sign(s) - \varepsilon_2|s|^\alpha sign|\dot{s}|) \tag{4.16}$$

$$\dot{V} = s\dot{s} - k_1\dot{s}^2 - k_2 s\dot{s} - \varepsilon_1|s|^\alpha \dot{s}.sign(s) - \varepsilon_2|s|^\alpha \dot{s}.sign|\dot{s}|) \tag{4.17}$$

$$\dot{V} \leq |\dot{s}|(s - k_1\dot{s} - k_2 s - \varepsilon_2|s|^\alpha|\dot{s}|) \tag{4.18}$$

$$\dot{V} \leq 0 \tag{4.19}$$

as $\varepsilon_i > 0, k_i > 0$.

Hence the global asymptotic stability is guaranteed since the derivative of the Lyapunov function is negative definite.

## 4.5 Controller design

Consider an Inverted pendulum system, considered for the swing-up and stabilization case. Control to the system is applied by means of the force 'u' to the cart, shown in Figure 1. Applying the Lagrange's method, the systems nonlinear equation of motion obtained as

$$\ddot{\theta} = \frac{mgl\sin(\theta) - m^2 l^2 \cos(\theta)\sin(\theta)\dot{\theta}^2 + u.ml\cos(\theta)}{m^2 l^2 \cos^2(\theta) - (I + ml^2)} \tag{4.20}$$

Consider $x_1$ and $x_2$ as pendulum angle and pendulum velocity respectively. Therefore,

$$\begin{cases} \dot{x}_1 = x_2 \\ \dot{x}_2 = \frac{mgl\sin x_1 - m^2 l^2 \cos x_1 \sin x_1 x_2^2 + u.ml\cos x_1}{m^2 l^2 \cos^2 x_1 - (I + ml^2)} \end{cases} \tag{4.21}$$

The main aim is to balance the pendulum upwards to a position desirable for the experiment and to maintain the position without letting it fall. The cart is driven by the dc motor, controlled by a controller. A disturbance force is applied on the top of the position.

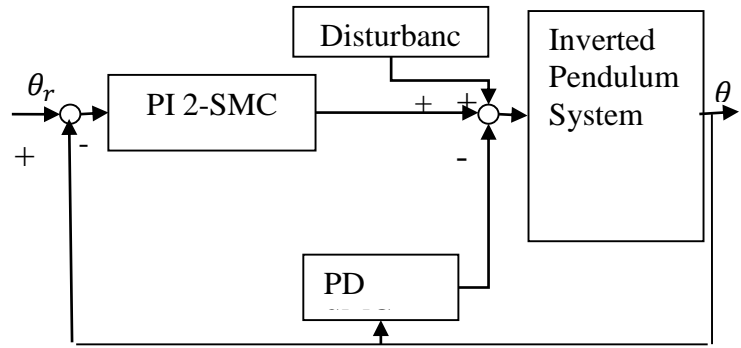

**Figure 4.1** Block diagram of Inverted Pendulum System with Sliding mode controller

## 4.6 Result and Discussion

According to the proposed algorithm, Parameters used are: $K_p$=2, $K_i$=125, $K_1$=125, $K_2$=95, $\eta$=5, c=25. Assuming that the model uncertainties dynamics is $0.5\sin(t)$ and external disturbance is $d(t) = 10\sin(t)$. For swing-up case, initial value is 180 degrees and for trajectory tracking state initial value is $[x_1, x_2] = [\pi\ 0]$ and the desired trajectory $x_d = 0.1\sin(t)$.

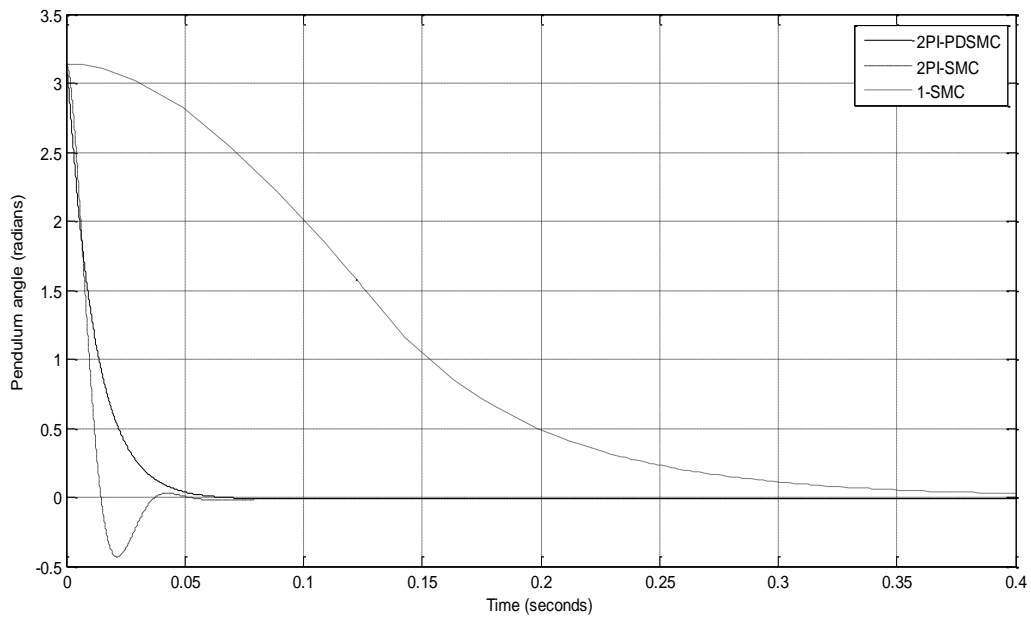

**Figure 4.2** Output (transient) response for different sliding mode control

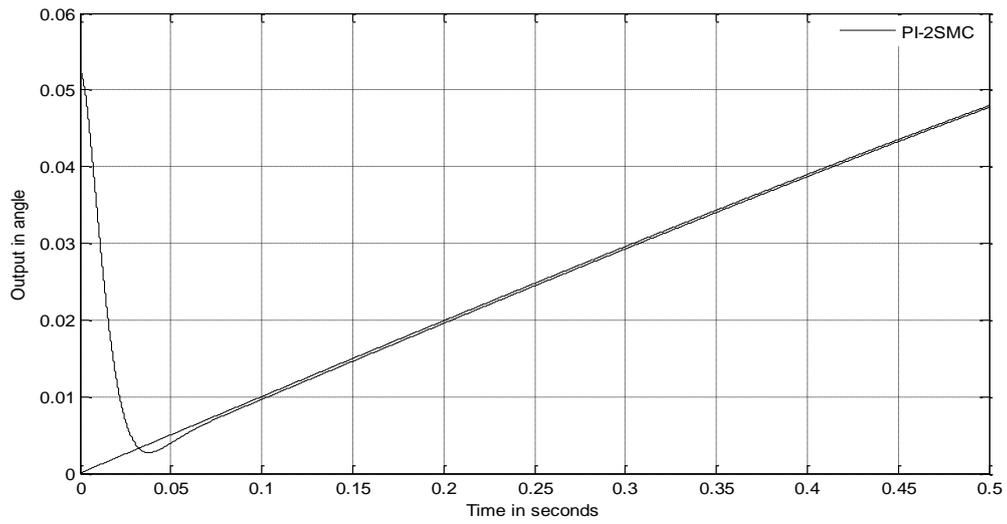

**Figure 4.3**. Trajectory tracking transient response curve.

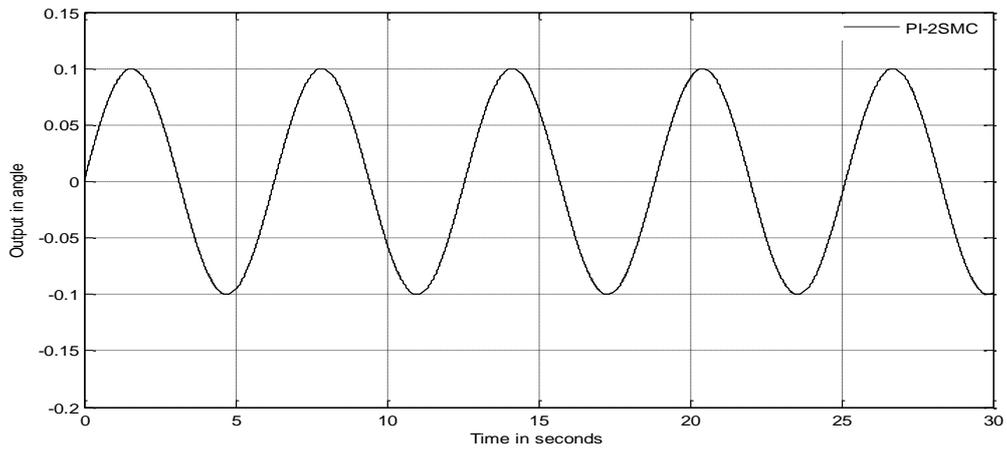

**Figure 4.4** Trajectory tracking response curve for the desired curve

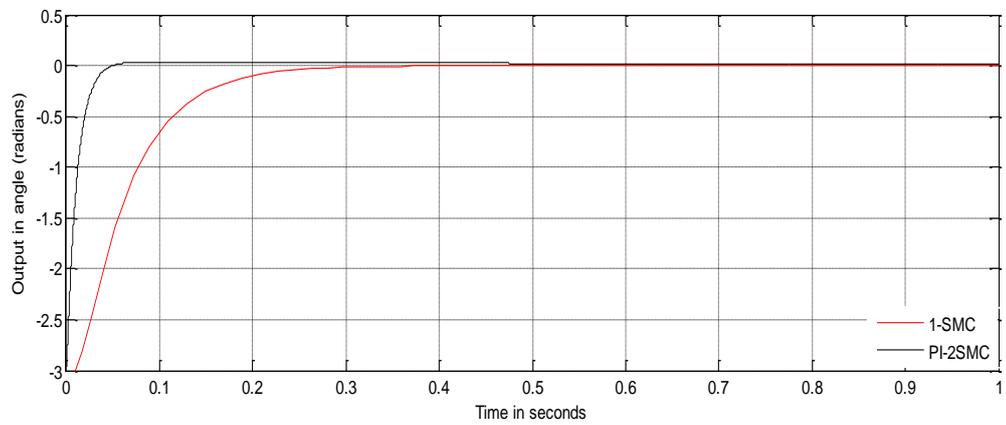

**Figure 4.5** Error Curve.

The output responses for swing up and stabilization of inverted pendulum are compared in the hierarchical manner in Figure 4.2. The transient response and steady state response for the trajectory tracking are shown in Figure 4.3 and 4.4. Error curve is shown in Figure 4.5. Simulations results show the controller is able to achieve desired characteristics successfully. Another point is that higher the order less is the chattering. Also the output has robustness to initial error and parametric uncertainties.

**Table 4.1**
**Simulation results of the Inverted Pendulum system**

| S.No. | Parameters | 1-SMC | PI-2SMC | PI-2SMC+PDSMC |
|---|---|---|---|---|
| 1 | Rise time (seconds) | 0.35 | 0.05 | 0.05 |
| 2 | Peak overshoot (radians) | 0 | 0.43 | 0 |
| 3 | Settling time (seconds) | 0.47 | 0.07 | 0.07 |
| 4 | Singularity | Yes | Yes | No |

## 4.7 Conclusion

PI (Proportional Integral) type sliding surface based second order sliding mode controller along the feed forward path with the negatively feed backed PD-type sliding surface based sliding mode controller, is designed for the swing up and stabilization of the Inverted Pendulum system with external disturbance. The proposed method nulls the singularity as well as increases the convergence time, sharply reducing the peak overshoot response. Also the improved reaching law accelerates the reaching velocity and improves the robustness of the dynamic process. The parameters are chosen appropriately to satisfy the desired requirement. This method solves the problem of high frequency chattering and also results in tracking of the desired trajectory accurately. Simulations results show the desired output response. It is observed from simulation results that, the response improves for the modified controller structure, which shows the effectiveness of the algorithm applied on the modified controller structure scheme.

# Chapter 5

# EXPERIMENTAL VALIDATION

## 5.1 Introduction

Technological world has been marked by learning new and improved methods to control the environment. Simply stated the term control means methods to force parameters in the environment to have specific values. In general, all the elements necessary to accomplish the control objective are described by the term control system. Different control strategies have been proposed in the previous chapters. In this chapter, the experimental validation of the proposed method is narrated.

A real time system is used which includes *Interacting / Non Interacting Conical Tank System*. Now days many research works has been carried out for such conical tank systems. In this experiment, single input single output (SISO) non-linear system is analyzed. The experiments are performed for a single tank system in a non-interacting mode. Since the tank is conical in shape, the variation of level of water is non-linear.

## 5.2 Conical Tank System

Conventional controllers are widely used in industries since they are simple, robust and familiar to the field operator. Practical systems are not precisely linear but may be represented as linearized models around a nominal operating point. The controller parameters tuned at that point may not reflect the real-time system characteristics due to variations in the process parameters. The solution is that the controller parameters have to be continuously adjusted. Figure 5.1 shows the block diagram conical tank.

### 5.2.1 Description and Working

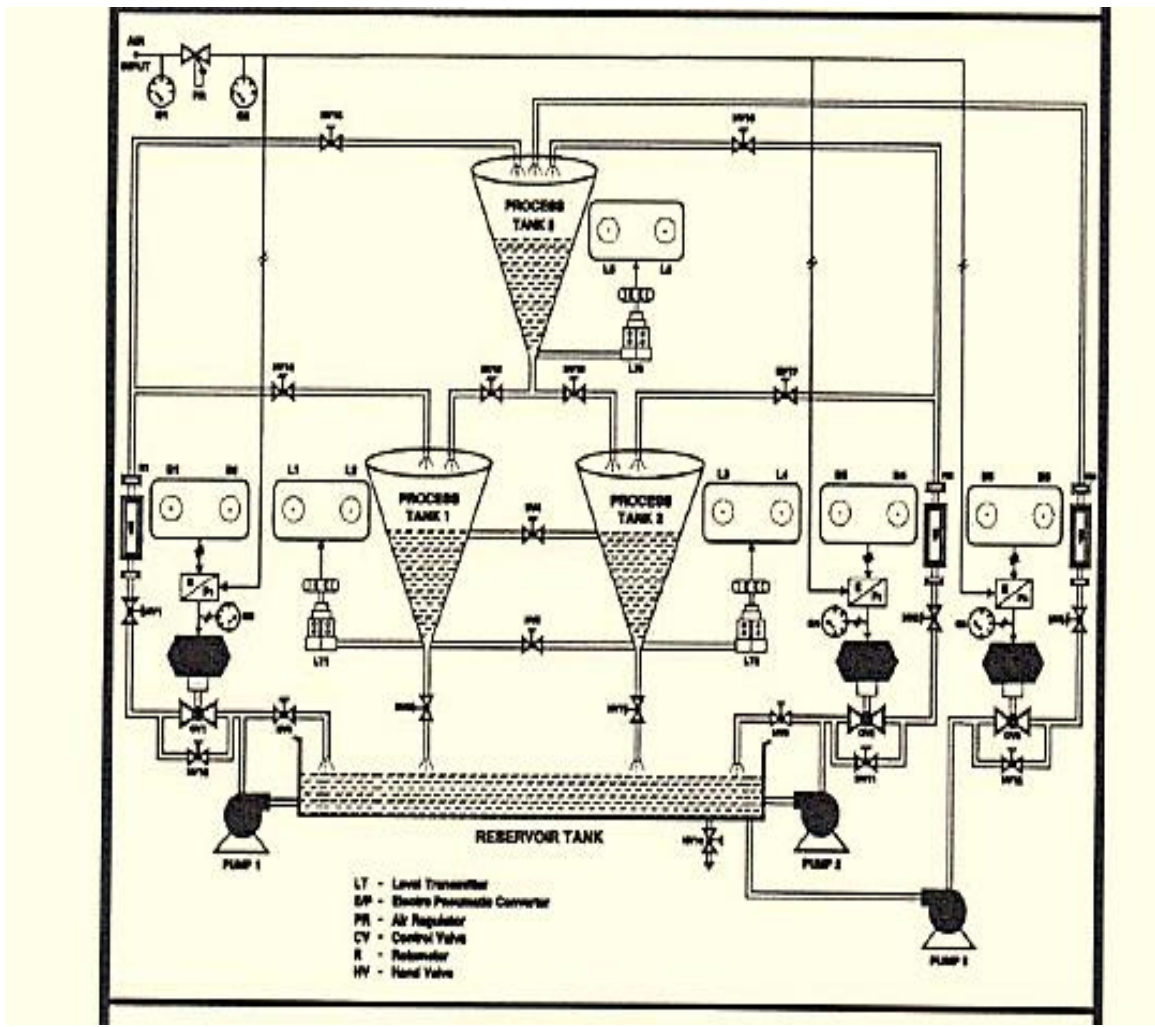

**Figure 5.1:** Front panel diagram of conical tank system

Water from storage tank is pumped with a discharging rate of 1500 LPH, continuously to the Stainless Steel (SS) conical tank of height 700 mm through a pneumatic control valve, whose valve action is air to open type. The DPLT transmits a current signal (output range 4-20mA) with a supply range of 24V DC @ 200mA to the I/V converter. The output of the I/V converter (1-5V) is given to the PC with VDPID-03 interfacing hardware consisting of multifunction high speed ADC and DAC on either side. The onboard data converters of the VDPID-03 can be directly linked with the Simulink tool of MATLAB thus forming a complete closed loop system. The signal from the PC is transmitted to the I/P converter through V/I converter. The output from the I/P converter which operates with a constant pressure of 20 psi, is pressured air in the range of 3-15 psi for actuating the control valve, which regulates the flow of liquid into the conical tank.

*Differential Pressure Transmitter (DPT)*

Differential pressure transmitter is works on the principle of force balance. DPT is used to measure the differential pressure (gauge pressure or absolute pressure). The output in terms of 4 - 20mA DC and it can be transmitted by a lead to other devices of controller.

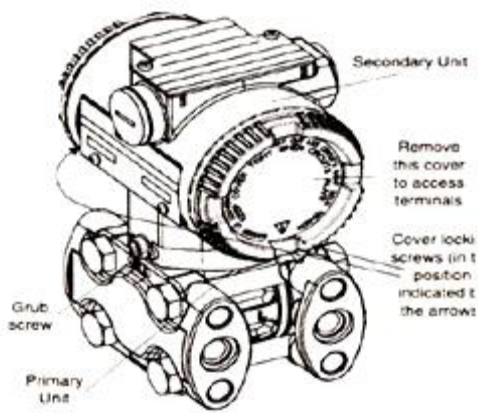
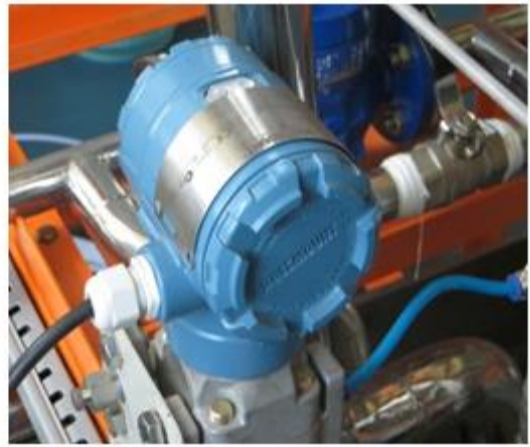

(i) Pictorial View          (ii) Real View

**Figure 5.2:** DPT Sensor

Differential Pressure Transmitter (DPT) is shown in Figure 5.2 which has primary sensor of diaphragm and secondary transducers of piezo electric sensor, primary and secondary a capsule type and piezo electrode sensor is a quartz crystal type.

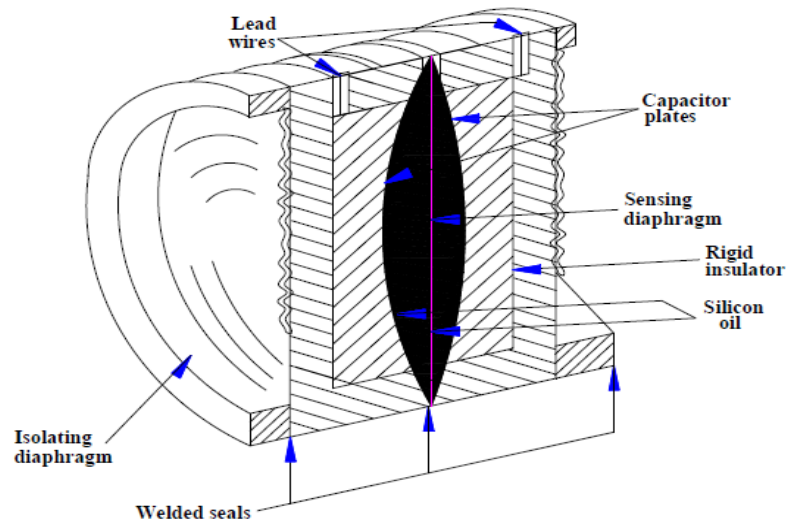

**Figure 5.3:** Internal Diagram

Capsule type of diaphragm is kept inside of the transmitter which is shown in Figure 5.3. Diaphragm movement depends upon the pressure difference between the terminals. The variation is allowed to strike one face of the crystal material, the crystal produces electrical energy by the principle of piezo electric effect. This output can be conditioned by using microcontroller based calibration technique.

**Technical Specification**

Type -           DPT
Source -         Rosemount
Supply Range -   24V DC @ 200mA
Measuring Range - (0-4000) mmWc
Output Range -   (4-20) mA / 2-wire system

*Current to Pressure Converter (I-P)*

Consider the arrangement as shown in Figure 5.4. The input current flows through the coil (1), thereby magnetizing the soft-iron yoke (2). The flux lines of this system being exposed at the gap (3) apply a force proportional to the input signal on the permanent magnet (4) which is made from a highly coercive metal. The small magnet (4) together with the flapper (5) forms the moving parts, controlling the air pressure at the nozzle (6), which is proportional to the magnetic force. The air flowing from the nozzle forms a restoring force balanced by the force applied to the magnet. The nozzle (6) is supplied with air through a throttle and back pressure through power amplifier gives proportional output. The described units are properly matched. Hence, a linear correspondence of electric input and pneumatic output signals is achieved. The direction of action of the converter is determined by the coil polarization. Zero adjustment is performed using the potentiometer connected with a resistor in parallel to the coil (1).

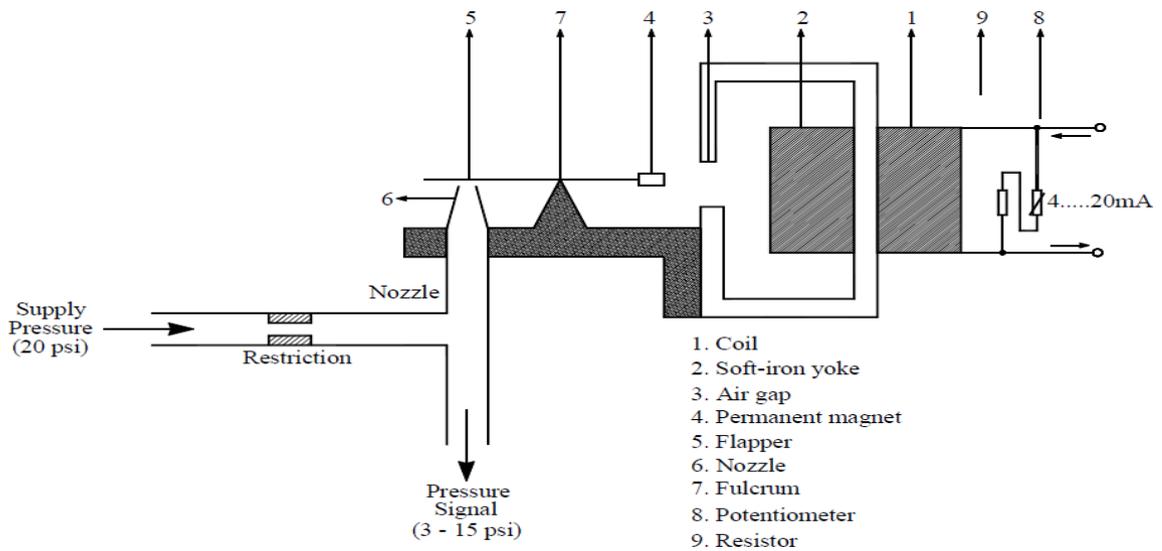

**Figure 5.4:** Arrangement for I-P conversion

The electro pneumatic (I/P) signal converter is used as a linking component between electric or electronic and pneumatic systems. It converts standard electric signals (4-20) mA, respectively into the standard pneumatic signal (3 - 15) psi. Due to its innovative construction principle based on a fixed coil and a low-mass (100 mg) moving permanent magnet, the I/P signal converter is highly resistant to shocks and vibration.

**Technical specification**

| | |
|---|---|
| Make - | ABB (Asea Brown Boveri Ltd.) |
| Input Air - | 20 psi constant pressure |
| Signal input - | (4 to 20) mA DC @ 20 psi |
| Output - | Pneumatic signal (3 to 15) psi |

*Pneumatic Control Valve*

The most common final control element is the pneumatic valve. This is an air-operated valve which controls the flow through an orifice by positioning approximately a plug. The plug is attached at the end of a stem which is supported on a diaphragm at the other end. As the air pressure (controller output) above the diaphragm increase, tends to moves down and consequently the plug restricts the flow through the orifice. Such a valve is known is an "air-to-close" valve.

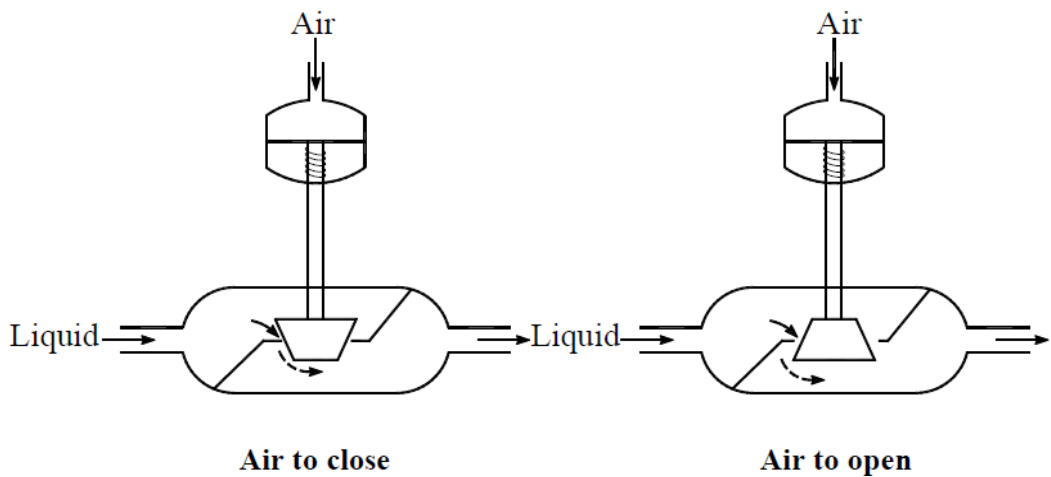

**Figure 5.5**: Operation of control valve

If the air supply above the diaphragm is lost, the valve will "fail open" since the spring would push the stem and the plug upward. There are pneumatic valves with opposite actions, (i.e. "air-to-open" which "fail closed") as shown in Figure 5.5. The most commercial valves move from fully closed as the air pressure at the top of the diaphragm changes from 3 to 15 Psi. There are three types of control valve flow characteristics namely on-off, linear and equal percentage. Figure 5.6 shows the change in flow rate with respect to the stem position for the three configurations.

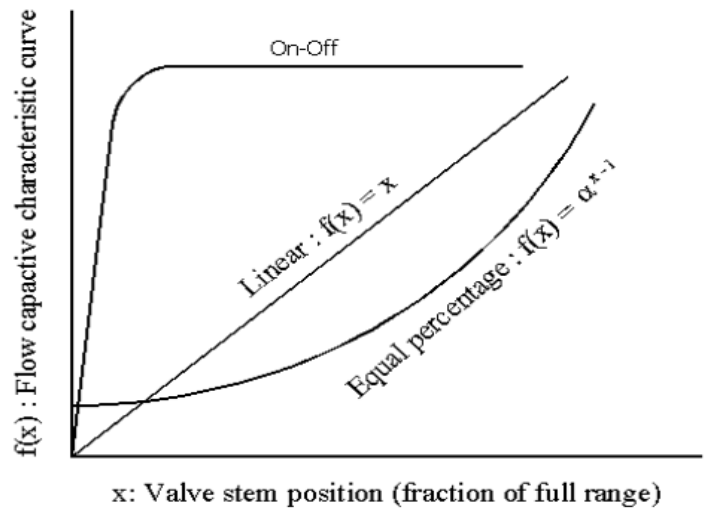

**Figure 5.6:** Variation of flow rate with stem position

**Calibration**

The pneumatic control valve is also known as adjustable area orifice. Based on pneumatic input (3-15) psi, control valve stem tends to move upward and downward thereby control the flow of the fluid through the pipe line. Since it is an air to open type, therefore

- At 15 psi pneumatic input, diaphragm of the valve tends to move upwards. It lifts the stem to fully open condition, and
- At 3 psi pneumatic input, diaphragm of the valve tends to move down wards. It pushes the stem to fully close condition.

**Technical Specification**

| | |
|---|---|
| Source - | RK valve ltd |
| Type - | Globe valve |
| Flow rate - | (500/1000) Liters / Hour |
| Trim mat - | SS316 |
| Characteristics - | Equal % |
| Spring range - | (0-2.1) Kg/cm$_2$ |
| Valve action - | Air to open |
| End connection - | 3/4" flange type |
| Medium - | Water / Air |
| Body Material - | CS body |

## 5.2.2 Experimental Results

In this section, the proposed controller is validated in a conical Tank system as shown in Figure 5.7. *Matlab 7.10.0 / Simulink* is used for experimental purpose. Initially, the process is modeled. Then, the controller is designed using proposed control design method. For modeling, the conical tank is represented as shown in Figure 5.8.

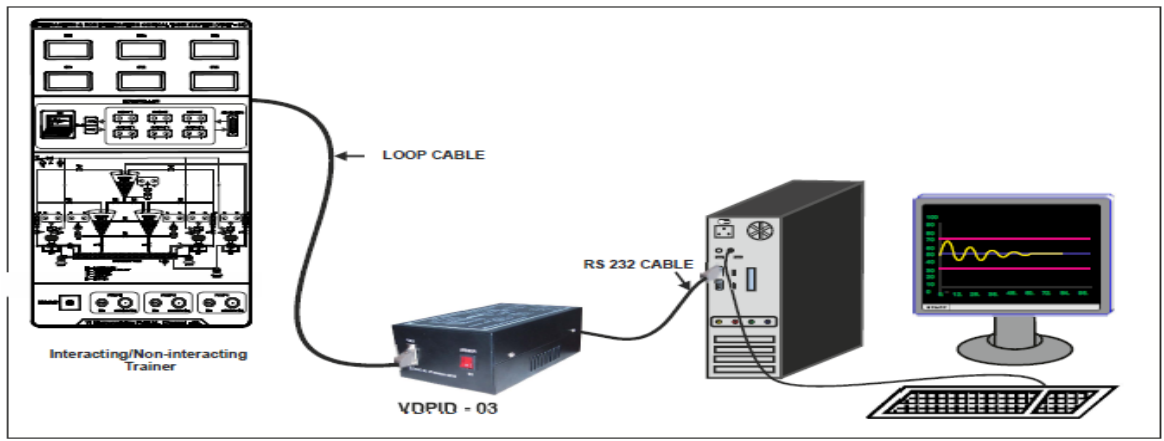

**Figure 5.7:** Interfacing diagram for controlling the set point through desktop

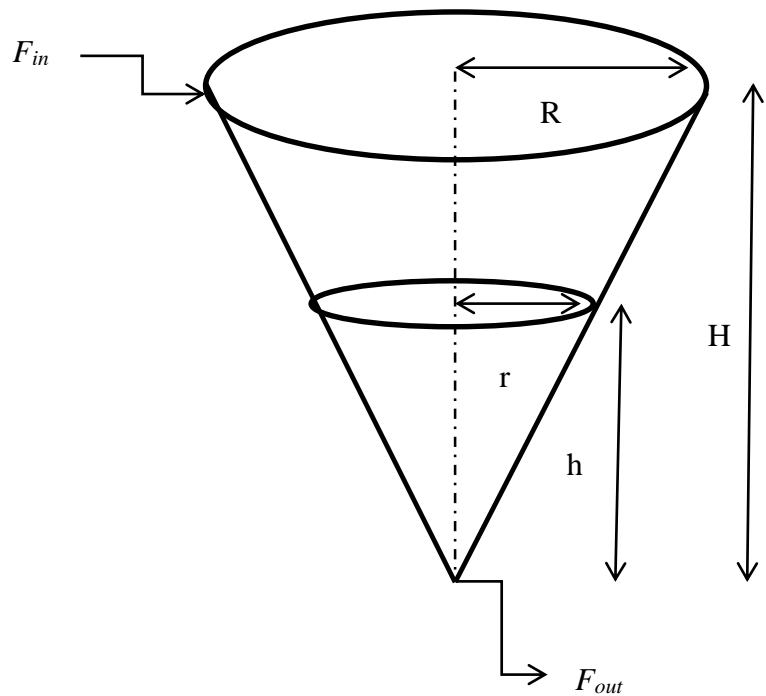

**Figure 5.8:** Structure of conical tank

Conical tank model is a benchmark problem in nonlinear control systems. In this level process, conical shape of the tank with constant changing area of cross section makes the system nonlinear. The objective is to maintain the level of the liquid at a desired level, which is achieved by controlling the input flow into the tank. The controlling variable is the level of the tank and the manipulated variable is the inflow to the tank. The liquid level will flow into the tank through inlet and the liquid will come out of the tank through outlet.

Feedback control system is designed based on the actuating error signal, which is the difference between the feedback signal and the set point, is fed to the controller to bring the output to the desired value. Error is fed to the controller for the proposed PID sliding surface design, converges the error to zero in finite time. The proposed sliding surface design will force the trajectory to converge to faster than the standard sliding mode, as will be discussed in simulations section.

The structure of the conical tank system is shown in Fig. 5.8. The tank level process to be simulated is a single input single output (SISO) tank system. The user can control the inflow rate by adjusting the control signal, Fin. During the simulations, the level 'h' will be calculated at any instant of time, t.

Dynamic model of conical tank is given as:

Area of the conical tank is given by:

$$A = \pi r^2 \tag{5.1}$$

$$\tan\theta = r/h = R/H \tag{5.2}$$

$$r = R * h/H \tag{5.3}$$

According to the law of conservation of mass,

Inflow rate – outflow rate = Accumulation.

$$F_{in} - F_{out} = A \frac{dh}{dt} \tag{5.4}$$

$$F_{out} = k\sqrt{h} \quad \text{where, k = discharge coefficient} \tag{5.5}$$

Therefore,

$$F_{in} - k\sqrt{h} = A \frac{dh}{dt} \quad \text{where,} \ A = \pi * R^2 * h^2 / H^2 \tag{5.6}$$

Rate of change of height,

$$\frac{dh}{dt} = \left. F_{in} - k\sqrt{h} \middle/ (\pi * R^2 * h^2 / H^2) \right. \tag{5.7}$$

**Table 5.1**
**Operating Parameters for Conical tank system**

| S. No. | Parameters | Description of the Quantity | Value |
|---|---|---|---|
| 1. | R | Total radius of the cone | 17.5 cm |
| 2. | H | Maximum total height of the tank | 70 cm |
| 3. | K | Value of the coefficient | 55 $cm^2/s$ |
| 4. | Fin | Maximum inflow rate of the tank | 400 LPH |

Control effort required for tracking control of the desired level process is given as

$$u(t) = -\frac{1}{K_p g(h,t)} \left( K_i (h - h_d) + K_p f(h) - k_{sc} * sgn(s(x)) \right) \tag{5.8}$$

$$u(t) == -(K_p . g(h,t))^{-1} . (K_i (h(t) - h_d(t)) + K_p (f(h)) + ks + k_{sc}|s|^\alpha . sat(s)) \tag{5.9}$$

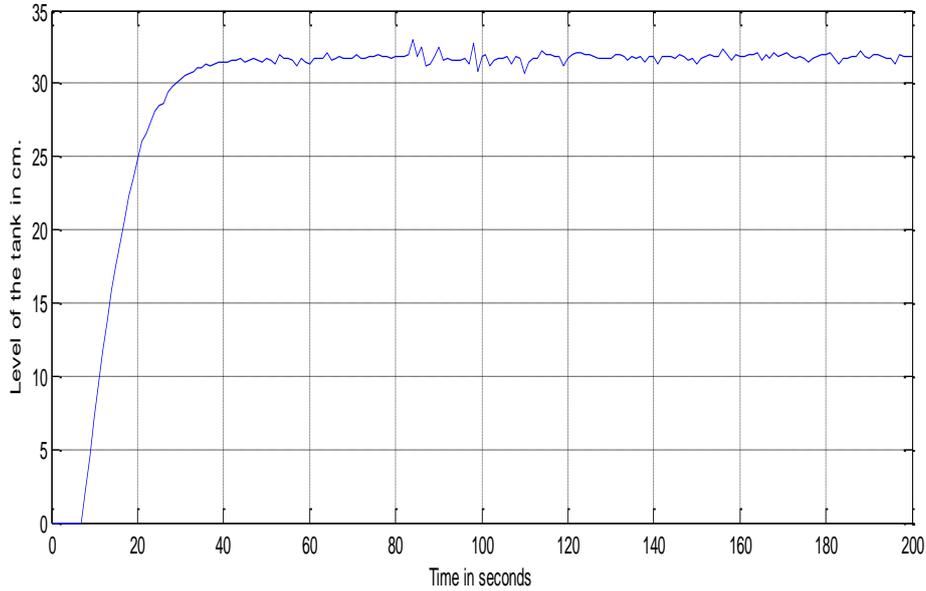

**Figure 5.9** Output with uncertainty 'd' for MPSO PID-SMC with improved reaching law.

## 5.3 Conclusion

The schematic block diagram is shown in Figure 5.1 with the parameters specified in Table 5.1. The system consists of a conical water tank, a water pump mechanism, a liquid level sensor and a PC-base controller. Dynamic equation of the conical tank is defined in [5.7] with the control effort given by [5.8]. The MPSO PID-SMC is proposed to be applied to the liquid level controller of a conical tank system. The objective is to control the liquid level of the tank by introducing a leakage (external disturbance) in the tank. The objective consists in minimizing the liquid level tracking error in presence of model uncertainties and leakage in the tank, limited only by the sensors capability to detect the level. The parameters for the proposed MPSO PID-SMC are chosen as $K_d = 0.8, K_i = 4.2, K_p = 105, k = 35 \text{ and } k_{sc} = 1.5$.

# Chapter 6

# Conclusion and Future Scope

## 6.1 Conclusion

The thesis has proposed modified forms of sliding mode controller structure using variants of PID-type sliding surfaces and modified forms of reaching laws for non-linear systems. Presence and forms of non-linearity in nature has been discussed which shows the prime importance of considering the non-linearities directly while analyzing and designing controllers for non-linear systems. Moreover using PSO as the offline optimization technique to tune the parameters for non-linear systems was discussed.

Initially, traditional sliding mode controller was designed using PSO algorithm for non-linear dynamical system and it was found that the traditional controller gives improved result both in settling time and disturbance than the traditional PID controllers. The control structure is then modified by adding an integral term in the sliding surface function which in turn modifies the controller, parameters of which are tuned using PSO technique. It is observed that the additional integral term in the controller structure improves the convergence time as well as the stability of the non-linear process that leads improvement in the performance of the controller. Also, the non-linear system is invariant to parameters uncertainties and external disturbances. The problem of chattering is encountered mainly in sliding mode control, reduced mainly by using the boundary conditions. Progressing on the problem of chattering, second order sliding mode controller based on PI-type sliding surface connected in the feed-forward path. Proposed is the technique where PD-type sliding mode control is connected in the feedback loop, reducing peak overshoot completely as well as the problem of singularity encountered in the SimMechanics model of the nonlinear inverted pendulum system during simulations. The stability of the non-linear system is proves using the direct Lyapunov stability criteria. The proposed controller is checked and verified using both simulation and experiment. Further, the robustness of the controller is verified by taking ±10% variation in the non-linear plant parameters and constant external disturbance connected to the plants. For experimental validation a real time conical system was analyzed, whose shape justifies its non-linearity using MatlabR2010a / Simulink. It is found that the experimental result is proving the proposed method by a high convergence rate. It shows satisfactory performance i.e. the settling time of the experimental result gives out 29 seconds.

## 6.2 Future Scope

Every real dynamical system shows visible nonlinear characteristics. These controllers are validated for SISO systems (Single stage nonlinear inverted pendulum). More complex systems can also be analyzed like Cart-Pendulum underactuated system (Multi Input Multi Output systems (MIMO)) involving the non-linearity. The controller leaves some work relating to the attenuation of the chattering characteristics by designing Observers. The use of higher order sliding mode control technique as well as higher order differentiators help to solve lot of advance system's control problem. Also, use of terminal sliding mode concept and backstepping technique provides some insight into reduced convergence characteristic and higher stability characteristics.

# Publications